\documentclass[10pt]{article}
\usepackage{graphicx}
\usepackage{amsfonts}
\usepackage{amsmath}
\usepackage{amssymb}
\usepackage{url}

\usepackage{dsfont}
\usepackage{fancyhdr}
\usepackage{indentfirst}
\usepackage{enumerate}
\usepackage[normalem]{ulem}
\usepackage{mathrsfs}
\usepackage{multirow}
\usepackage{diagbox}
\usepackage{bm}
\usepackage{todonotes,soul}
 \usepackage[colorlinks=true,citecolor=blue]{hyperref}
\usepackage{amsthm}
\usepackage{color}

\definecolor{DarkGreen}{rgb}{0.2,0.6,0.2}

\def\blue#1{\textcolor{blue}{#1}}

\definecolor{purple}{rgb}{0.6,0.3,0.8}

\usepackage{natbib}
\usepackage{comment}

\def\d{\mathrm{d}}

\newcommand{\E}{\mathbb{E}}

\newcommand{\R}{\mathbb{R}}

\newcommand{\p}{\mathbb{P}}
\newcommand{\X}{\mathcal{X}}

\renewcommand{\H}{\mathcal{H}}

\newcommand{\Q}{\mathbb{Q}}

\newcommand{\id}{\mathds{1}}

\renewcommand{\ge}{\geqslant}
\renewcommand{\le}{\leqslant}
\renewcommand{\geq}{\geqslant}
\renewcommand{\leq}{\leqslant}
\renewcommand{\epsilon}{\varepsilon}

\renewcommand{\cdots}{\dots}
\theoremstyle{plain}
\newtheorem{theorem}{Theorem}
\newtheorem{corollary}{Corollary}

\newtheorem{proposition}{Proposition}
\theoremstyle{definition}

\newtheorem{example}{Example}

\theoremstyle{remark}
\newtheorem{remark}{Remark}

\DeclareMathOperator*{\argmin}{arg\,min}

\newcommand{\dsquare}{\mathop{  \square} \displaylimits}
\newcommand{\VaR}{\mathrm{VaR}}

\newcommand{\ES}{\mathrm{ES}}

\newcommand{\trd}{\textcolor{black}}

\newcommand{\thedate}{\today}

\DeclareMathOperator{\sign}{sign}

\usepackage{setspace}

\usepackage{footmisc}
\begin{document}
\title{Lambda Value-at-Risk under ambiguity and risk sharing}
\author{
Peng Liu\thanks{School of Mathematics, Statistics and Actuarial Science, University of Essex, UK. Email: peng.liu@essex.ac.uk} \and Alexander Schied\thanks{Department of Statistics and Actuarial Science,
  University of Waterloo,  Canada. Email: aschied@uwaterloo.ca}
}

  \date{\thedate}

 \maketitle
\begin{abstract}
In this paper, we investigate the Lambda Value-at-Risk ($\Lambda\VaR$) under ambiguity, where the ambiguity is represented by a family of probability measures. We establish that for increasing Lambda functions, the robust (i.e., worst-case) $\Lambda\VaR$ under such an ambiguity set is equivalent to $\Lambda\VaR$ computed with respect to a capacity, a novel extension in the literature. This framework unifies and extends both traditional $\Lambda\VaR$ and Choquet quantiles (Value-at-Risk under ambiguity). 
We analyze the fundamental properties of this extended risk measure and establish a novel equivalent representation for $\Lambda\VaR$ under capacities with monotone Lambda functions in terms of families of downsets. Moreover, explicit formulas are derived for robust $\Lambda\VaR$ when ambiguity sets are characterized by  $\phi$-divergence and the likelihood ratio constraints, respectively.

We further explore the applications in risk sharing among multiple agents. We demonstrate that the family of risk measures induced by families of downsets is closed under inf-convolution. In particular, we prove that the inf-convolution of  $\Lambda\VaR$ with capacities and monotone Lambda functions is  another $\Lambda\VaR$ under a capacity.    The explicit forms of optimal allocations are also derived. Moreover,  we obtain more explicit results for risk sharing under ambiguity sets characterized by $\phi$-divergence and likelihood ratio constraints.  Finally, we  explore comonotonic risk-sharing for $\Lambda\VaR$ under ambiguity.

\begin{bfseries}Key-words\end{bfseries}: Lambda Value-at-Risk under capacities; Value-at-Risk; Choquet quantiles; Risk sharing; Inf-convolution; Model uncertainty; $\phi$-divergence; Kullback-Leibler divergence; Likelihood ratio
\end{abstract}

 \section{Introduction}\label{sec:1}
\nocite{*}
Model uncertainty, also known as ambiguity in decision theory, plays a central role in decision-making and risk quantification. The Basel regulatory framework (\cite{BASEL19}) explicitly requires that risks need to be assessed under a variety of stressed probability scenarios. This introduces ambiguity into the evaluation process, as considering multiple probability models reflects underlying model uncertainty. A widely used paradigm to address this is the application of worst-case risk measures across these scenarios, which is a well-established approach in modeling decisions under ambiguity. It was first axiomatized by \cite{GS89} through the maxmin expected utility framework, and subsequently extended by \cite{HS01} (multiplier preferences), \cite{GMM04} (the $\alpha$-maxmin model), \cite{MMR06} (variational preferences), and \cite{CHMM21} (preferences under model misspecification). 
Worst-case risk measures under multiple probability distributions have been extensively studied in the context of risk optimization under uncertainty; see \cite{GOO03}, \cite{NPS08}, and \cite{ZF09} for applications to Value-at-Risk (VaR) and Expected Shortfall (ES).

Recently,  it \trd{was} shown in \cite{LMW24} that the worst-case quantiles under a set of probability measures correspond to the so-called Choquet quantiles, introduced in \cite{C07}. These quantiles are defined with respect to a capacity--a non-additive prior. Choquet quantiles belong to the broader framework of decision-making under non-additive probabilities, as studied in \cite{Y87} (dual utility theory), \cite{Q82} (rank-dependent expected utility), and \cite{S89} (Choquet expected utility). The risk sharing with Choquet quantiles was also studied in \cite{LMW24}.

Inspired by \cite{LMW24}, we investigate the worst-case of Lambda Value-at-Risk ($\Lambda\VaR$) under a set of probability measures and study the risk sharing problem for multiple $\Lambda\VaR$ under model uncertainty. The target is to find the explicit expression of  the worst case of $\Lambda\VaR$ similarly as Choquet quantiles, and the expressions of the inf-convolution and the optimal allocations for risk sharing under model uncertainty.   Differently from the setup in this paper, the robust model of $\Lambda\VaR$ under a set of distribution functions was established in \cite{HL24}.

In Section \ref{sec:properties}, we analyze the fundamental properties of $\Lambda\VaR$ defined with respect to capacities, denoted by $\Lambda\VaR^{w}$, where $w$ is a capacity and $\Lambda$ is a function. \blue{It has the following form:}
$$\Lambda\VaR^w(X)=\inf\{x\in \R: w(X>x)\leq \Lambda(x) \}.$$ 
One essential property is that for increasing $\Lambda$, the $\Lambda\VaR^{w}$ is closed under the $\sup$ operator, i.e., the worst case of $\Lambda\VaR^{w}$ under a set of capacities is a $\Lambda\VaR^{w^*}$ with $w^*$ a new capacity. This result extends the result for Choquet quantiles established in \cite{LMW24}. If the ambiguity set consists of a set of probability measures, this property shows that the robust model for $\Lambda\VaR$ with increasing $\Lambda$ is $\Lambda\VaR^{w}$ with $w$ being a capacity. This is the motivation for us to study $\Lambda\VaR^{w}$. \trd{For general $\Lambda$,  the robust model of $\Lambda\VaR$ over a set of probability measures is also $\Lambda\VaR^{w}$ with $w$ a capacity in the sense of  model aggregation (MA)  with respect to the first order stochastic dominance, recently studied by \cite{MWW25}. Other properties  such as monotonicity, cash subadditivity and quasi-star-shapedness are also examined}.  Moreover, \trd{analogously to} \cite{HWWX24}, a representation of $\Lambda\VaR^w$ is also derived in terms of Choquet quantiles for increasing $\Lambda$. Finally, we establish a novel representation of $\Lambda\VaR^w$  in terms of  families of downsets.  We find that $\Lambda\VaR^w$ with monotone $\Lambda$ functions corresponds to the risk measures induced by monotone \trd{families} of downsets, offering an equivalent representation for  $\Lambda\VaR^w$ and playing an important role in the study of risk sharing in Section \ref{sec:multiple}.

In Section \ref{Sec:robustVaR}, we consider the robust $\Lambda\VaR$ with increasing $\Lambda$ under  two uncertainty sets: i) the uncertainty sets induced by $\phi$-divergence including the well-known Kullback-Leibler (KL) divergence and  chi-squared divergence as special cases; ii) the uncertainty sets defined by the likelihood ratio, where the Radon-Nikodym  derivative of the underlying probability measure with respect to a reference probability measure is bounded by two random variables from above and below respectively. Surprisingly, the robust $\Lambda\VaR$ under the first uncertainty set is equivalent to a $\Lambda\VaR$ under the reference probability with a transformed $\Lambda$ function. This conclusion also holds true for the second uncertainty set if the two random variables are constants. If the two random variables for the second uncertainty set are not constants, we also derive the robust models for $\Lambda\VaR$, and  a more explicit expression is derived for the case that the random variable controlling the lower bound is zero.

Section \ref{sec:multiple} is devoted to the problem of risk sharing  with multiple agents. We find that  the inf-convolution operator is closed for risk measures induced by families of downsets, which include all $\Lambda\VaR^w$ as special cases.  This implies that the inf-convolution of $\Lambda\VaR^w$ is a risk measure induced by a family of downsets. We also derive the corresponding optimal allocations by splitting and distributing the tail events among agents, in a manner  similarly to the model  without uncertainty considered in \cite{ELMW19}, \cite{L24} and \cite{XH24}. 
Interestingly, we further demonstrate that the inf-convolution of multiple $\Lambda\VaR^w$, when all $\Lambda$ functions are monotone in the same direction, results in another $\Lambda\VaR^w$. This extends the related results for Choquet quantiles in \cite{LMW24}, and suggests that the inf-convolution of multiple $\Lambda\VaR$ with increasing $\Lambda$ functions under ambiguity yields a $\Lambda\VaR^{w}$ with $w$ being a capacity.   In fact, our results are very general, and unify and extend the prior findings: the results for $\VaR$ in \cite{ELW18} and \cite{ELMW19} under both homogeneous and heterogeneous beliefs,  for $\VaR$ in \cite{LMW24} under model uncertainty, and for $\Lambda\VaR$ in \cite{L24}, \cite{XH24} and \cite{LSW24}. Finally, the two specific uncertainty sets introduced in Section \ref{Sec:robustVaR} is revisited in the context of risk sharing. Under the assumption that all $\Lambda$ functions are increasing, we derive more explicit expressions for the inf-convolution and the optimal allocations.

In Section \ref{Sec:Comonotonic}, the risk sharing for several $\Lambda\VaR^{w}$ is considered by assuming all allocations are comonotonic. This restriction makes the problem qualitatively different from the one in Section \ref{sec:multiple}. We find that, under some assumptions, the optimal allocation is to distribute the total risk to a single agent and  to distribute nothing to other agents. Moreover, we obtain some equivalent condition on the $\Lambda$ functions such that the previous conclusion holds true. The risk sharing problem under the two specific uncertainty sets introduced in Section \ref{Sec:robustVaR} is also discussed.

Notation and definitions are given in Section \ref{sec:2} and all the proofs are delegated to the Appendix.

\subsection{Related literature}
The $\Lambda\VaR$ was introduced by \cite{FMP14} as an extension of $\VaR$, enabling the use of multiple confidence levels through a loss-dependent function $\Lambda$. The choice of the $\Lambda$ function is highly flexible, as it can be either increasing or decreasing. The selection of $\Lambda$ based on data was explored in \cite{HMP18}. Compared to $\VaR$, one advantage of  $\Lambda\VaR$ is its ability to distinguish tail risk when $\Lambda$ is decreasing, in a manner similar to the Loss $\VaR$ proposed by \cite{BBM20}. For increasing $\Lambda$ functions, the $\Lambda\VaR$ can also incorporate additional elements, such as the judgment of risk managers in the risk management process; see \cite{BP22}.  Furthermore, recent literature confirms several desirable properties of the $\Lambda\VaR$. \cite{HWWX24} showed that, for an increasing $\Lambda$ function, the $\Lambda\VaR$ satisfies quasi-star-shapedness, a property weaker than quasi-convexity that helps mitigate risk concentration. Additionally, $\Lambda\VaR$ with increasing $\Lambda$ functions satisfies cash subadditivity, which is particularly useful for measuring future financial loss under stochastic interest rates; see \cite{KR09} and \cite{HWWX24}. For properties like monotonicity, locality, and others, we refer to \cite{BP22}, and for robustness, elicitability, and consistency, see \cite{BPR17}. \trd{The characterization of $\Lambda\VaR$ was given in \cite{BP22} and \cite{C05} respectively.}

The $\Lambda\VaR$ has been applied to several problems in quantitative risk management recently. The risk sharing problem for $\Lambda\VaR$ under homogeneous beliefs was studied in \cite{L24} and \cite{XH24}, and under heterogeneous beliefs it was studied in \cite{LSW24}, where expressions for the inf-convolution and forms of the optimal allocation were derived. Additionally, the $\Lambda\VaR$ has been applied to optimal reinsurance design with model homogeneity in \cite{BBB23} and \cite{BCHW24}, as well as to robust portfolio selection and sensitivity analysis in \cite{HL24} and \cite{IPP22}. 

Since the introduction of convex and coherent risk measures by \cite{ADEH99}, \cite{FS02} and \cite{FR05}, the risk sharing problem for risk measures has been studied extensively. The risk sharing for convex risk measures was studied by \cite{BE05}, \cite{JST08} and \cite{FS08}, showing the  optimal allocations are comonotonic if the risk measures are law-invariant.   Recently, more focus has been placed on the risk sharing problem for non-convex risk measures, such as quantile-based risk measures including Value-at-Risk ($\VaR$) and Expected Shortfall ($\ES$) as special cases (e.g., \cite{ELW18}, \cite{LMWW22} and \cite{W18}) and distortion riskmetrics including inter-quantile-range as an example (e.g., \cite{LLW23a}). In those papers, it is shown that the optimal allocation is pairwise countermonotonic, as defined in \cite{LLW23}. The risk sharing problem with heterogeneous beliefs was studied in  \cite{ELMW19}, \cite{L20}, \cite{L22}, \cite{LSW24}, \cite{AGP15} and \cite{BG20}. Under ambiguity, the risk sharing for several $\VaR$ was investigated in \cite{LMW24}, deriving the expressions of the inf-convolution and the optimal allocations, and the comonotonic risk sharing for $\VaR$ with uncertainty sets characterized by Wassertein ball was studied in \cite{LMWW22}.

\section{Notation and Definitions}\label{sec:2}

For a given measurable space $(\Omega,\mathcal F)$, let $\mathcal X_0$ denote the collection of all bounded random variables. 
Moreover, let $\mathcal X$ be a set of random variables containing  $\X_0$. We do not specify $\X$ but we  suppose $\X$ has good enough properties to conduct our study \trd{such as $\X=\X_0$ or $\X_1$, where $\X_1$ represents the set of all random variables}. 
  
  For a mapping $\rho:\X\to\R$, we say that $\rho$ is \emph{monotone} if $X\leq Y$ implies $\rho(X)\leq \rho(Y)$; and $\rho$ is \emph{cash-additive} if $\rho(X+c)=\rho(X)+c$ for $X\in\X$ and $c\in\R$;  $\rho$ is 
{\emph{cash-subadditive} (resp. cash-supadditive)} if $\rho(X+m)\leq\rho(X)+m$ (resp. $\rho(X+m)\geq\rho(X)+m$) for all $X\in \X$ and $m\ge 0$; $\rho$ is  {\emph{quasi-star-shaped}} if $\rho(\lambda X+(1-\lambda) t) \leqslant \max \{\rho(X), \rho(t)\}$  for all $X \in \mathcal{X}, t \in \mathbb{R}$ and $\lambda \in[0,1]$. We say $\rho$ is a \emph{monetary} risk measure if it is monotone and cash-additive.  We refer to Chapter 4 of \cite{FS16} for more details on the properties, characterization and applications of risk measures.

 In the context of capital requirements,  cash-subadditivity considers the time value of money by allowing a non-linear increase in the capital requirement as cash is added to the future financial position, and describes the situations with stochastic or ambiguous interest rate; see \cite{KR09} and \cite{HWWX24}. Cash-subadditivity is  satisfied by convex loss-based risk measures proposed in \cite{CDH13}.  Quasi-star-shapedness is a weak version of quasi-convexity, and its theoretic-decision interpretation can be found at  \cite{HWWX24}. 
  



Let us now define capacities and Lambda Value-at-Risk ($\Lambda\VaR$) with respect to capacities. We say that $w:\mathcal F\to[0,1]$ is \emph{increasing} if $A\subseteq B$ implies $w(A)\leq w(B)$; it is a \emph{capacity} if $w$ is increasing satisfying $w(\emptyset)=0$ and $w(\Omega)=1$; it is \emph{continuous} if $\lim_{n\to\infty}w(A_n)=0$ whenever $A_n\downarrow \emptyset$. \trd{In this paper, the terms ``increasing'' and ``decreasing'' are used in the non-strict sense.}

For $\Lambda:\R\to [0,1]$ and a capacity $w$, the $\Lambda\VaR$ \trd{with respect to  $w$} are given by
 \begin{align}\label{lambdaL}\Lambda\VaR^w(X)&=\inf\{x\in \R: w(X>x)\leq \Lambda(x) \},\nonumber\\
 \Lambda\VaR^{+,w}(X)&=\sup\{x\in \R: w(X>x)\geq \Lambda(x) \},
 \end{align}
where $\inf\emptyset=\infty$ and $\sup\emptyset=-\infty$.

\trd{Let $\hat{w}(A)=1-w(A^c)$ for $A\in\mathcal F$ and $\hat{\Lambda}(x)=1-\Lambda(x),~x\in\R$. Then $\hat{w}$ is a capacity and we further have the equivalent definitions as $\Lambda\VaR^w(X)=\inf\{x\in \R: \hat{w}(X\leq x)\geq \hat{\Lambda}(x) \}$ and
 $\Lambda\VaR^{+,w}(X)=\sup\{x\in \R: \hat{w}(X\leq x)\leq \hat{\Lambda}(x) \}$.  As the form of \eqref{lambdaL} is more convenient for risk sharing application, we use \eqref{lambdaL} throughout this paper.}

Let $\p$ be a probability measure on $(\Omega,\mathcal F)$. If $w=\p$, then $\Lambda\VaR^w$ and $\Lambda\VaR^{+,w}$ boil down to \trd{$\Lambda\VaR$ and $\Lambda\VaR^{+}$}  proposed in \cite{FMP14}; see \cite{BP22}, \cite{HWWX24} and \cite{C05} for the properties, representation and characterization  of \trd{the corresponding} Lambda Value-at-Risk.

  If $\Lambda$ is a constant, then $\Lambda\VaR^{w}$ and $ \Lambda\VaR^{+,w}$ boil down to Choquet quantiles, \trd{first} introduced by \cite{C07} as an extension of Value-at-Risk ($\VaR$) under a probability to the case under a capacity, i.e.,
$\VaR^{w}$ at level $p\in [0,1]$ is given by
 $$
\VaR_p^{w}(X) =\Lambda\VaR^{w}(X),~~~~X\in \mathcal  X,
 $$ for $\Lambda=p$. As it is shown in \cite{LMW24},  \trd{a Choquet quantile} is the robust version of $\VaR$ under a set of probability measures. Inspired by this,   we will show later that \trd{for increasing $\Lambda$}, $\Lambda\VaR^w$ and $\Lambda\VaR^{+,w}$ are the robust versions of $\Lambda\VaR$ under a set of probability measures.


   We denote by $\mathcal H$ the collection of all functions $\Lambda: \mathbb{R}\to (0,1)$, and $\mathcal H_I$ (resp. $\mathcal H_D$) the collection of all increasing (resp. decreasing) functions in $\H$.  Hereafter, for any $\Lambda\in \mathcal H$, let $\lambda^-=\inf_{x\in\R}\Lambda(x)$ and $\lambda^+=\sup_{x\in\R}\Lambda(x)$. We denote by $[n]$ the set $\{1,\dots,n\}$. 
   
Next, we \trd{recall the notion of} the inf-convolution. A random variable $X\in \mathcal{X}$ represents a risk to be shared between $n$ agents. Then, define the set of \emph{allocations} of $X$ as
 \begin{equation*}
\mathbb{A}_n(X)=\left\{(X_1,\ldots,X_n)\in \mathcal{X}^n: \sum_{i=1}^nX_i=X\right\}. \label{eq:intro1}
\end{equation*}
The \emph{inf-convolution} of the risk measures $\rho_1,\dots,\rho_n$ is the mapping  $\dsquare_{i=1}^n \rho_i:\X\to[-\infty,\infty)$, defined as
\begin{equation*}\label{eq:inf-con}
 \dsquare_{i=1}^n \rho_i (X)  = \inf\left\{\sum_{i=1}^n\rho_i(X_i): (X_1,\cdots,X_n)\in \mathbb A_n(X)  \right\}.
 \end{equation*}

An $n$-tuple $(X_1,\cdots,X_n)\in \mathbb A_n(X)$ is called \emph{an optimal allocation} of $X$ for $(\rho_1,\dots,\rho_n)$ if $\sum_{i=1}^{n}\rho_i(X_i)=\dsquare_{i=1}^n \rho_i(X)$.  The inf-convolution $\dsquare_{i=1}^n \rho_i (X)$ can be interpreted as  the smallest possible aggregate capital  or the smallest risk exposure for the total risk $X$ in the financial system with $n$ agents.  More economic interpretations on the inf-convolution can be found in e.g., \cite{D12}, \cite{R13} and \cite{ELW18}.

For the risk measures $(\rho_1,\dots,\rho_n)$ and a total risk $X$, an allocation $(X_1,\cdots,X_n)\in \mathbb A_n(X)$ is \emph{Pareto-optimal} if for any other allocation $(Y_1,\dots,Y_n)\in \mathbb A_n(X)$, $\rho_i(Y_i)\leq \rho_i(X_i)$,  $i \in [n]$ imply $\rho_i(Y_i)=\rho_i(X_i)$,  $i \in [n]$.   For  finite-valued monetary risk measures, it is shown in \cite{ELW18} that an allocation  is optimal if and only if it is Pareto-optimal. \trd{In Section \ref{sec:properties}}, we will show that $\Lambda\VaR^w$  does not satisfy cash-additivity in general, implying that it is not a monetary risk measure. However, one can still show that the optimal allocation of the inf-convolution of $\Lambda\VaR^w$ is Pareto-optimal.

  \section{Properties and representations of $\Lambda\VaR^{w}$}\label{sec:properties}
In this section, we explore the properties of $\Lambda\VaR^{w}$ and $\Lambda\VaR^{+,w}$ and find \trd{some} alternative representations. Some of them are crucial in the risk sharing applications in Section \ref{sec:multiple}. We first list some basic properties.
  \begin{proposition}\label{Prop:properties} For a capacity $w$ and $\Lambda, \Lambda_1, \Lambda_2\in\H$, $\Lambda\VaR^{w}$ and $\Lambda\VaR^{+,w}$ satisfy the following properties.
  \begin{enumerate}[(i)]
      \item $\Lambda\VaR^{w}$ and $\Lambda\VaR^{+,w}$ are monotone and $\Lambda\VaR^{w}\leq \Lambda\VaR^{+,w}$;
      \item $\Lambda_1\VaR^w\geq \Lambda_2\VaR^w$ and $\Lambda_1\VaR^{+,w}\geq \Lambda_2\VaR^{+,w}$ if $\Lambda_1\leq \Lambda_2$;
      \item $\Lambda\VaR^{w}$ and $\Lambda\VaR^{+,w}$ satisfy quasi-star-shapedness if $\Lambda$ is increasing;
      \item $\Lambda\VaR^{w}$ and $\Lambda\VaR^{+,w}$ satisfy cash-subadditivity  if $\Lambda$ is increasing and cash-supadditivity if $\Lambda$ is decreasing.
  \end{enumerate}
  \end{proposition}
For a class  $W$ of capacities, we write 
\begin{equation}
\label{eq:max-w} \overline w =\sup_{w\in W} w~\text{and}~\underline{w} =\inf_{w\in W} w.
\end{equation}
   
\begin{proposition}\label{prop:motive}
    For $\Lambda\in \H_I$  and a class of capacities $W$, we have
    $$\sup_{w\in W}\Lambda\VaR^{w}=\Lambda\VaR^{\overline w}~\text{and}~\inf_{w\in W}\Lambda\VaR^{+,w}=\Lambda\VaR^{+,\underline{w}}.$$
\end{proposition} 

    Proposition \ref{prop:motive} includes the following important special cases, which provide one of the motivations to study {$\Lambda\VaR$ with respect to a capacity}.
    
   \begin{corollary}\label{Cor:motive}  For $\Lambda\in \H_I$ and  a class of probabilities $\mathcal P$, we have
   $$\sup_{\mathbb Q\in \mathcal P}\Lambda\VaR^{\mathbb Q}=\Lambda\VaR^{\overline {\mathbb Q}}~\text{and}~\inf_{\mathbb Q\in \mathcal P}\Lambda\VaR^{+,\mathbb Q}=\Lambda\VaR^{+,\underline{\mathbb Q}},$$
   where $\overline {\mathbb Q}=\sup_{\mathbb Q\in \mathcal P}\mathbb Q$ and $\underline {\mathbb Q}=\inf_{\mathbb Q\in \mathcal P}\mathbb Q$ are capacities. 
   \end{corollary}

For decreasing $\Lambda\in \H$, the conclusion in Proposition \ref{prop:motive} fails to hold in general.  \trd{This can be seen from} the following example. \trd{For $x,y\in\R$, let $x\vee y=\max(x,y)$ and $x\wedge y=\min(x,y)$.}
\begin{example}\label{Exam:1} Let $\Lambda(x)=((1-x)\vee 0.2)\wedge 0.8,~x\in\R$ and $\mathbb P_1$ and $\mathbb P_2$ live on $([0,1],\mathcal B([0,1]))$, where $\mathcal B([0,1])$ is the set of all Borel measurable sets on $[0,1]$. Let $X(q)=(3/4)\id_{[0,5/8]}(q)+(1/2)\id_{[5/8,3/4]}(q)$ for $q\in [0,1]$.
We suppose $\mathbb P_1$ is the Lebesgue measure and $\mathbb P_2$ is a probability measure satisfying $\mathbb P_2((3/4,1])<0.2$ and $\mathbb P_2([0,5/8])<0.2$. 
Direct computation shows \trd{$\Lambda\VaR^{\mathbb P_1}(X)=0$} and $\Lambda\VaR^{\mathbb P_2}(X)=1/2$. Moreover, for $\overline{\mathbb P}=\mathbb P_1\vee\mathbb P_2$, we have  $\Lambda\VaR^{\overline{\p}}(X)=3/4$. Hence, $\Lambda\VaR^{\mathbb P_1}(X)\vee \Lambda\VaR^{\mathbb P_2}(X)<\Lambda\VaR^{\overline{\p}}(X)$. This shows that the conclusion  in Proposition \ref{prop:motive} does not hold for decreasing $\Lambda$.  We can similarly construct counterexamples for $\Lambda\VaR^{+,w}$. 
\end{example}

{Proposition \ref{prop:motive} and Example \ref{Exam:1} suggest that the case of an increasing function $\Lambda$ is the most interesting one.} 
\trd{For general $\Lambda$ and and  a class of probabilities $\mathcal P$, if $X$ is bounded from above, then $\Lambda\VaR^{\overline {\mathbb Q}}(X)$ with $\overline {\mathbb Q}=\sup_{\mathbb Q\in \mathcal P}\mathbb Q$ is the robust evaluation of $\Lambda\VaR^{\mathbb Q}$ over $\mathcal P$ via  model aggregation (MA)  with respect to the first order stochastic dominance, recently studied by \cite{MWW25}, which follows directly from Proposition 1 of \cite{MWW25}.  Hence, the case of general function $\Lambda$ is also meaningful. In the sequel,  we will sometimes also state results for more general $\Lambda$, provided that the corresponding statements can be obtained without too much effort. }

\begin{proposition}\label{prop:CQ} For  $\Lambda\in \H_I$ and a capacity $w$, we have
   $$\Lambda\VaR^{w}(X)=\inf_{x\in\R} \left\{\VaR_{\Lambda(x)}^{w} (X) \vee x\right\}=\sup_{x\in \mathbb{R}} \left\{\VaR_{\Lambda(x)}^{w} (X) \wedge x\right\}$$
   and 
   $$\Lambda\VaR^{+,w}(X)=\inf_{x\in\R} \left\{\VaR_{\Lambda(x)}^{+,w} (X) \vee x\right\}=\sup_{x\in \mathbb{R}} \left\{\VaR_{\Lambda(x)}^{+,w} (X) \wedge x\right\}.$$
\end{proposition}

Proposition \ref{prop:CQ} shows that $\Lambda\VaR^{w}$ and $\Lambda\VaR^{+,w}$ can be rewritten as the infimum or supremum of some Choquet quantiles, extending Theorem 3.1 of \cite{HWWX24}, where the case $w=\p$ was considered. It is worth mentioning that in Proposition \ref{prop:CQ}, $\Lambda\VaR^{w}$ and $\Lambda\VaR^{+,w}$ may take {the} values $\pm\infty$, as there is no requirement on the continuity of $w$ and on the range of $\Lambda$ with $0<\lambda^-\leq \lambda^+<1$.  {We believe that the representations obtained in Proposition \ref{prop:CQ} can be useful in applications to optimal reinsurance design, as investigated in \cite{BCHW24} for the case $w=\p$}.

Next, we give an equivalent representation  of $\Lambda\VaR^w$ with monotone $\Lambda$. We say $\mathcal A\subseteq \mathcal F$ is a \emph{downset}\footnote{The terminology here is consistent with \cite{C07} but different from the one in \cite{LMW24}, where it is called \emph{null set} \trd{of a capacity}. } if $\emptyset\in \mathcal A$ 
and $A\in \mathcal A$ whenever $B\in\mathcal A$ and $A\subseteq B$.
In \cite{C07}, a Choquet quantile can be represented in terms of a downset. However, for $\Lambda\VaR^w$, a single downset is not enough. Instead, we need a family of downsets. 

For $\Lambda\in \H$ and a capacity $w$, let \begin{align}\label{Nullseq}\mathcal A_x=\{A\in\mathcal F: w(A)\leq \Lambda(x)\},~x\in\R.
\end{align}
 Note that $\{\mathcal A_x\}_{x\in\R}$ is a \trd{family} of downsets.
 Moreover, $\Lambda\VaR^w$ can be recovered through this family of downsets as below:
\begin{align*}\Lambda\VaR^w(X)=\inf\{x\in\R:\{X>x\}\in \mathcal A_x\},~X\in\X.
\end{align*}
Clearly, any $\Lambda\VaR^w$ corresponds to a family of downsets defined by \eqref{Nullseq}. Conversely, for a family of downsets $\{\mathcal A_x\}_{x\in\R}$,  we define 
\begin{align}\label{Eq:Recover}\rho(X)=\inf\{x\in\R:\{X>x\}\in \mathcal A_x\},~X\in\X.
\end{align} 
We can easily check that $\rho$ satisfies monotonicity. We say $\rho$ defined by \eqref{Eq:Recover} is a \emph{risk measure induced by a family of downsets} $\{\mathcal A_x\}_{x\in\R}$. For simplicity, we call $\rho$ an \emph{induced risk measure}.  Clearly, $\Lambda\VaR^w$ belongs to the family of induced risk measures.  
The question is whether an induced risk measures  is always \trd{of the form} $\Lambda\VaR^w$ for some $\Lambda\in\mathcal H$ and a capacity $w$.

 To simplify our analysis, we only focus on the monotone family of downsets defined as follows. For a family of sets $\{\mathcal A_x\}_{x\in\R}$, we say it is \emph{increasing} (resp. \emph{decreasing}) if $\mathcal A_x\subseteq \mathcal A_y$ (resp. $\mathcal A_x\supseteq \mathcal A_y$) whenever $x\leq y$. For a decreasing family $\{\mathcal A_x\}_{x\in\R}$, we say it is  \emph{left-continuous} if  $\bigcap_{y<x}\mathcal A_y=\mathcal A_x$  for all $x\in\R$.

Let $\mathcal H_I^*$ (resp. $\mathcal H_{D}^*$) denote all the increasing (resp. decreasing) functions $\Lambda:\R\to (0,1]$. Clearly, $\mathcal H_I\subsetneq \mathcal H_I^*$ and $\mathcal H_D\subsetneq \mathcal H_D^*$. The extension of the value of $\Lambda$ to including $1$ is \trd{needed} for the application in risk sharing in Section \ref{sec:multiple}. 
Note that if $\Lambda\in \mathcal H_I^*$, then $\{\mathcal A_x\}_{x\in\R}$ defined by \eqref{Nullseq} is increasing; if $\Lambda\in \mathcal H_D^*$ is left-continuous, then $\{\mathcal A_x\}_{x\in\R}$ defined by \eqref{Nullseq} is decreasing and left-continuous. 
Surprisingly, the converse conclusion also holds true; see the following theorem. This means that $\Lambda\VaR^w$ with monotone $\Lambda$ functions and the monotone families of downsets can be mutually determined in a way similar to the Choquet quantiles; see \cite{C07} and \cite{LMW24}. 
\begin{theorem}\label{Pro:representation} 
Suppose $\rho:\mathcal X\to[-\infty,\infty]$.
\begin{enumerate}[(i)]
\item   The risk measure $\rho$ is of the form  $\Lambda\VaR^w$ for some $\Lambda\in \mathcal H_I^*$ and  a capacity $w$ if and only if  there exists an increasing  family of downsets $\{\mathcal A_x\}_{x\in\R}$ such that \eqref{Eq:Recover} holds.
 \item The risk measure $\rho$ is of the form  $\Lambda\VaR^w$ for some left-continuous $\Lambda\in \mathcal H_D^*$ and  a capacity $w$ if and only if  there exists a decreasing and left-continuous family of downsets $\{\mathcal A_x\}_{x\in\R}$ such that \eqref{Eq:Recover} holds.
    \end{enumerate}
\end{theorem}

Note that Theorem \ref{Pro:representation} offers an equivalent representation of  $\Lambda\VaR^w$ with monotone $\Lambda$ functions via \eqref{Eq:Recover}. This new representation will be very helpful in the risk sharing problem studied in Section \ref{sec:multiple}. If $\mathcal A_x=\mathcal A_y$ for all $x,y\in\R$, then \eqref{Eq:Recover} recover the result for Choquet quantiles given in \cite{C07}.  If $\Lambda$ and $w$ are fixed, then the downsets can be constructed through \eqref{Nullseq}. \trd{Conversely, however, constructing $\Lambda$ and $w$ from a given family of downsets is more complicated}. As it is shown in the proof of Theorem \ref{Pro:representation} in Appendix \ref{Appendix:Sec3},  $\Lambda$ can be constructed by \eqref{Eq:ReLambda} for an increasing family of downsets and  the corresponding $w$ can be given by  $w(A)=\Lambda(x)$ if $A\in \mathcal A_x\setminus \mathcal A_x^-$ for $x\in\R$; $w(A)=\frac{\Lambda(x)+\Lambda(x+)}{2}$ if $A\in \mathcal A_x^+\setminus \mathcal A_x$ for $x\in\R$; $w(A)=0$ if $A\in \bigcap_{x\in\R} \mathcal A_x$; $w(A)=1$ if $A\in \mathcal F \setminus\left(\bigcup_{x\in\R} \mathcal A_x\right)$ with $\mathcal A_x^+=\bigcap_{y>x}\mathcal A_y$ and $\mathcal A_x^-=\bigcup_{y<x}\mathcal A_y.$

Compared with part (i) of Theorem \ref{Pro:representation}, \trd{part (ii) of Theorem} \ref{Pro:representation} additionally requires the left continuity for the family of downsets. This is due to the complex structure of $\Lambda\VaR^w$ with decreasing $\Lambda$ functions. The conclusion may not be correct if  this assumption is removed. Fortunately, this additional \trd{assumption still allows} for the application to the risk sharing in Section \ref{sec:multiple}.

The risk measures induced by increasing  families of downsets  satisfy the following property.
\begin{proposition}\label{Prop:downsets} Suppose $\{\mathcal A_x\}_{x\in\R}$ is an increasing  family of downsets and $\{\mathcal B_x\}_{x\in\R}$ satisfies $\mathcal B_x\subseteq \bigcap_{y>x}\mathcal A_y$ and $\mathcal B_x\supseteq \bigcup_{y<x}\mathcal A_y$ for $x\in\R$.  For $X\in\X$, we have
\begin{align*}\inf\left\{x\in\R:\{X>x\}\in\mathcal B_x \right\}=\inf\{x\in\R:\{X>x\}\in \mathcal A_x\}.
\end{align*}
    
\end{proposition}

Proposition \ref{Prop:downsets} means that for an increasing family of downsets, a slight change on those sets does not affect the risk measures defined by  \eqref{Eq:Recover}. \trd{This result} becomes complicated for decreasing family of downsets and the conclusion may not be correct.

\section{Robust $\Lambda\VaR$}\label{Sec:robustVaR}
In light of Corollary \ref{Cor:motive}, for $\Lambda\in\mathcal H_I$, the robust models for $\Lambda\VaR$ heavily rely on  $\overline {\mathbb Q}=\sup_{\mathbb Q\in \mathcal P}\mathbb P$ and $\underline {\mathbb Q}=\inf_{\mathbb Q\in \mathcal P}\mathbb Q$.  In this section, we seek the explicit expression of $\overline {\mathbb Q}$ and $\underline {\mathbb Q}$ under two different ambiguity sets $\mathcal P$. Note that $$\underline {\mathbb Q}(A)=1-\overline {\mathbb Q}(A^c).$$
Hence, we only need to find $\overline {\mathbb Q}$.
\subsection{Uncertainty sets induced by $\phi$-divergence}
For a convex $\phi:[0,\infty)\to\R\cup\{\infty\}$ satisfying $\phi(1)=0$ and $\phi(0)=\lim_{x\downarrow 0}\phi(x)$, and $\Q\ll\p$, the $\phi$-divergence is defined as 
$$D_\phi(\Q||\p)=\E^\p\left(\phi\left(\frac{\d \Q}{\d\p}\right)\right).$$
In particular, if $\phi(x)=(x-1)^2$, then the $\phi$-divergence is the so-called chi-squared divergence; if $\phi(x)=x^{\alpha}-1$ with $\alpha\geq 1$, it corresponds to the $\alpha$-divergence; if $\phi(x)=x\ln(x)$, it is the well-known Kullback-Leibler (KL) divergence; see e.g., \cite{GS02}, \cite{A09} and \cite{BenTal13}.

The uncertainty sets induced by $\phi$-divergence are defined  by $$\mathcal P_\phi(\p,\delta)=\left\{\mathbb Q\ll\mathbb P: D_\phi(\Q||\p)\leq \delta \right\},~\delta>0.$$


\begin{proposition}\label{Prop:alpha} For $\overline{\mathbb Q}(A)=\sup_{\mathbb Q\in \mathcal P_\phi(\p,\delta)}\mathbb Q(A)$, we have 
    $$\overline{\mathbb Q}(A)=g_{\phi,\delta}(\p(A)),~A\in\mathcal F,$$
    where $$g_{\phi,\delta}(x)=\sup\{t\in [x,1]: x\phi(t/x)+(1-x)\phi((1-t)/(1-x))\leq \delta\},~x\in (0,1),$$ $g_{\phi,\delta}(0)=0$ and $g_{\phi,\delta}(1)=1$. Moreover, if $\phi:[0,\infty)\to\R$ is strictly convex and $\lim_{x\to\infty}\frac{\phi(x)}{x}=\infty$, then $g_{\phi,\delta}$ is continuous and strictly increasing over $[0,x_\delta]$  and $g_{\phi,\delta}(x)=1$ over $[x_{\delta},1]$ with $x_\delta\in (0,1)$ being the solution to $x\phi(1/x)+(1-x)\phi(0)=\delta$; for $x\in (0,1)$, $g_{\phi,\delta}(x)>x$ and  $ g_{\phi,\delta}(x)\downarrow x$ as $\delta\downarrow 0$.
\end{proposition}

Next, we see some specific examples. Clearly, if $\phi$ is strictly convex and $\lim_{x\to\infty}\frac{\phi(x)}{x}=\infty$, then  $g_{\phi,\delta}(x)=1$ for $x\in [x_{\delta},1]$ and $g_{\phi,\delta}(0)=0$. In what follows, we only focus on $g_{\phi,\delta}(x)$ for $x\in (0,x_\delta)$ for the first three examples.
\begin{example}\label{Example}
    \begin{enumerate}[(i)]
    \item If $\phi(x)=x\ln(x)$, then  $x_{\delta}=e^{-\delta}$ and $g_{\phi,\delta}(x)\in (x,1)$ is the solution to $t\ln(t/x)+(1-t)\ln ((1-t)/(1-x))=\delta$  for $x\in (0,x_\delta)$;
    \item If $\phi(x)=x^{\alpha}-1$ with $\alpha>1$, then  $x_{\delta}=(1+\delta)^{1/(1-\alpha)}$  and $g_{\phi,\delta}(x)\in (x,1)$ is the solution to $t^\alpha x^{1-\alpha}+(1-t)^{\alpha}(1-x)^{1-\alpha}=1+\delta$  for $x\in (0,x_\delta)$;
     \item If $\phi(x)=(x-1)^2$, then  $x_{\delta}=1/(1+\delta)$ and $g_{\phi,\delta}(x)=x+\sqrt{\delta x(1-x)}$ for $x\in (0,x_\delta)$;
     \item \trd{If $\phi(x)=\infty\id_{[0,k_1)\cup (k_2,\infty)}(x)$ for some $0\leq k_1<1<k_2<\infty$, then $g_{\phi,\delta}(x)=(k_{2}x)\wedge (k_{1}x+1-k_{1})$ for $x\in [0,1]$.}
    \end{enumerate}
\end{example}

{The following theorem shows that robustifying $\Lambda\VaR^\Q$ by taking the supremum over $\Q\in\mathcal P_\phi(\p,\delta)$ is equivalent to using a standard Lambda $\VaR$ with respect to $\p$ but with a distortion of the original function~$\Lambda$.}

\begin{theorem}\label{cor:upper1} {If $\delta>0$, then } for $\Lambda\in\H_I$,  
$${\sup_{\Q\in \mathcal P_\phi(\p,\delta)}\Lambda\VaR^\Q=\Lambda\VaR^{g_{\phi,\delta}\circ\p},}$$
and for $\alpha\in (0,1)$, 
$${\sup_{\Q\in \mathcal P_\phi(\p,\delta)}\VaR_{\alpha}^\Q=\VaR^{g_{\phi,\delta}\circ\p}_\alpha.}$$
Moreover, if $\phi:[0,\infty)\to\R$ is strictly convex and  $\lim_{x\to\infty}\frac{\phi(x)}{x}=\infty$, then 
$${\sup_{\Q\in \mathcal P_\phi(\p,\delta)}\Lambda\VaR^\Q=(g_{\phi,\delta}^{-1}\circ\Lambda)\VaR^\p, ~\sup_{\Q\in \mathcal P_\phi(\p,\delta)}\VaR_{\alpha}^\Q=\VaR_{g_{\phi,\delta}^{-1}(\alpha)}^\p,}$$
and $$\lim_{\delta\downarrow 0}\sup_{\Q\in \mathcal P_\phi(\p,\delta)}\Lambda\VaR^\Q=\lim_{\delta\downarrow 0}(g_{\phi,\delta}^{-1}\circ\Lambda)\VaR^\p=\Lambda\VaR^{+,\p}(X).$$
\end{theorem}

The conclusion in the above corollary shows that the robust $\Lambda\VaR$ under the uncertainty set induced by $\phi$-divergence  is still a $\Lambda\VaR$ under the reference probability $\p$ with a transformed Lambda function, and the robust $\VaR$ under this uncertainty set is still a $\VaR$ under the reference probability $\p$ with a transformed confidence level. Hence, the uncertainty set induced by the $\phi$-divergence does not change the nature of $\Lambda\VaR$ and $\VaR$. However, the robust models of $\Lambda\VaR$  may not converge to $\Lambda\VaR$  under the reference probability $\p$ as the uncertainty parameter $\delta\downarrow 0$ if $\Lambda\VaR^{+,\p}(X)\neq \Lambda\VaR^{\p}(X)$.

For an increasing {function} $g:[0,1]\to [0,1]$ with $g(0)=0$ and $g(1)=1$, the distortion risk measure $\rho_g:\X\to\R$ is defined by
$$\rho_g^\Q(X)=\int_0^\infty g(\Q(X>x))\d x+\int_{-\infty}^0 (g(\Q(X>x))-1)\d x.$$
Distortion risk measures form a popular class of risk measures applied in insurance pricing, performance evaluation, decision theory and many other topics; see e.g., \cite{FS16}, \cite{Wang96}, \cite{Wang00}, \cite{CM09} and \cite{Y87}. It includes {the} two regulatory risk measures $\VaR$ and expected shortfall ($\ES$) as special cases.

Applying Proposition \ref{Prop:alpha} to distortion risk measures, we arrive at the following upper bound for robust distortion risk measures.
\begin{corollary}\label{cor:upper}  For $X\in \X$, we have 
    $$\sup_{\mathbb Q\in \mathcal P_\phi(\p,\delta)}\rho_g^{\Q}(X)\leq \rho_{g\circ g_{\phi,\delta}}^\p(X).$$
    Moreover, if $\phi$ is strictly convex, $\lim_{x\to\infty}\frac{\phi(x)}{x}=\infty$, and  $g$ is right-continuous over $[0,1)$, then $$\lim_{\delta\downarrow 0}\sup_{\mathbb Q\in \mathcal P_\phi(\p,\delta)}\rho_g^{\Q}(X)=\lim_{\delta\downarrow 0}\rho_{g\circ g_{\phi,\delta}}^\p(X)=\rho_{g}^\p(X).$$
\end{corollary}

In Corollary \ref{cor:upper}, we offer a simple upper bound for $\sup_{\mathbb Q\in \mathcal P_\phi(\p,\delta)}\rho_g^{\Q}(X)$. This upper bound is  close to the real value if the uncertainty ($\delta$) is small and $g$ is right-continuous. Hence, this simple upper bound may be useful in practice to quantify the risk under model uncertainty.

    
    \subsection{Uncertainty set controlled by the likelihood ratio} 
For an atomless probability measures $\mathbb P$,  let
\begin{align}\label{Eq:likelihood}
\mathcal P(\p,Y_1,Y_2)=\left\{\mathbb Q\ll\mathbb P: Y_1\leq \frac{d\mathbb Q}{d\mathbb P}\leq Y_2\right\},
\end{align}
where $Y_1\leq Y_2$ are two  nonnegative random {variables} satisfying $\mathbb E^{\mathbb P}(Y_1)<1<\mathbb E^{\mathbb P}(Y_2)<\infty$. 

Note that the likelihood ratio in the above uncertainty set is controlled by two random variables, which is different from the one used in \cite{LMWW22} and \cite{LLS24}, where the case of $Y_1=0$ and $Y_2=k>1$ was considered. {Note that the case $Y_1=k_1$ and $Y_2=k_2$ for two constants  $0\leq k_1<1<k_2<\infty$ was considered in part (iv) of our Example \ref{Example}.}

\begin{proposition}\label{thm:likelihood} For the uncertainty set defined by \eqref{Eq:likelihood}, we have the following conclusion.
For $A\in\mathcal F$, if  $\mathbb E^{\mathbb P}(Y_2\id_{A}+Y_1\id_{A^c})\leq 1$, then
    $$\overline{\mathbb Q}(A)=\mathbb E^{\mathbb P}(Y_2\id_{A});$$  
    If $\mathbb E^{\mathbb P}(Y_2\id_{A}+Y_1\id_{A^c})>1$, then 
    $$\overline{\mathbb Q}(A)=\mathbb E^{\mathbb P}(Y_2\id_{B}+Y_1\id_{A\setminus B}),$$
    where $B$ is \trd{any measurable subset of $A$} such that $\mathbb E^{\mathbb P}(Y_2\id_{B}+Y_1\id_{B^c})=1$.
\end{proposition}

Note that if $\mathbb E^{\mathbb P}(Y_2\id_{A}+Y_1\id_{A^c})>1$, then the set $B\subset A$ satisfying  $\mathbb E^{\mathbb P}(Y_2\id_{B}+Y_1\id_{B^c})=1$ may not be unique.  This is clear when $Y_1$ and $Y_2$ are constants.

In case of  $Y_1=k_1$ and $Y_2=k_2$ with $0\leq k_1<1<k_2$, the uncertainty set defined in \eqref{Eq:likelihood} implies that the discrepancy  of likelihood of the event has bounded ratio in the sense that $k_{1}\p(A)\leq \Q(A)\leq k_{2}\p(A)$ for all $A\in\mathcal F$. In case of $Y_1=0$, then there is no lower bound requirement on the likelihood ratio. In both two cases, $\overline{\mathbb Q}$ has explicit expression.
\begin{proposition}\label{Prop:likelihood1} Suppose the uncertainty set is given by \eqref{Eq:likelihood}.
\begin{enumerate}[(i)] 
\item If $Y_1=0$, then $\overline{\mathbb Q}(A)=1\wedge \mathbb E^{\mathbb P}(Y_2\id_{A})$ for all $A\in\mathcal F$, and $\sup_{\Q\in\mathcal P(\p,0,Y_2)}\Lambda\VaR^\Q=\Lambda\VaR^{\overline{\mathbb Q}}$;
\item  If $Y_1=k_1$ and $Y_2=k_2$, then 
$$\overline{\Q}(A)=g(\p(A)),~A\in\mathcal F,~\sup_{\Q\in\mathcal P(\p,k_1,k_2)}\Lambda\VaR^\Q=(g^{-1}\circ\Lambda)\VaR^{\mathbb P},$$
where
$g(x)=(k_{2}x)\wedge (k_{1}x+1-k_{1})$ for $x\in [0,1]$.
\end{enumerate}
\end{proposition}

Note that the above proposition shows that the robust $\Lambda\VaR$ under $\mathcal P(\p,k_1,k_2)$ is still a $\Lambda\VaR$ under $\p$ with a transformed $\Lambda$ function, which extends the results  in \cite{LMWW22} and \cite{XH24}, where only  the case $k_1=0$ was considered. However, the robust $\Lambda\VaR$ under $\mathcal P(\p,0,Y)$ is a $\Lambda\VaR$ under a capacity, not a probability measure anymore in general. This shows the necessity  of our study of $\Lambda\VaR^w$ with $w$ being capacities.

\section{Inf-convolution of multiple $\Lambda\VaR^{w}$}\label{sec:multiple}
 In this section, we first consider the {problem of} risk sharing for $\Lambda\VaR^{w}$. {To this end, we first consider the case of general risk measures that are induced by families of downsets. Then, using Theorem \ref{Pro:representation}, we move on to the particular case of $\Lambda\VaR^{w}$.}
  In the second part, we consider  risk sharing for multiple $\Lambda\VaR$ under model uncertainty with increasing $\Lambda$  by focusing on  two specific uncertainty sets: the one induced by $\phi$-divergence and the one defined by the likelihood ratio.   We aim to find the expressions for the inf-convolution and also the corresponding optimal allocations.
 
\subsection{Inf-convolution of $\Lambda\VaR^w$}\label{sec:multiple_general}
For {given} families of {downsets} $\{\mathcal A_x^{(i)}\}_{x\in\R},~i\in [n]$, define $\rho_i: \X\to[-\infty,\infty]$ as  
\begin{align}\label{Eq:rhoi}\rho_i(X)=\inf\{x\in\R:\{X>x\}\in \mathcal A_x^{(i)}\},~X\in\X.
\end{align}
Moreover, let $$\widetilde{\mathcal A}_x=\bigcup_{y_1+\dots+y_n=x}\oplus_{i=1}^n\mathcal A_{y_i}^{(i)},~\text{with}~\oplus_{i=1}^n\mathcal A_{y_i}^{(i)}=\{\cup_{i=1}^n A_i: A_i\in \mathcal A_{y_i}^{(i)},~i\in [n]\}.$$
Clearly, $\{\widetilde{\mathcal A}_x\}_{x\in\R}$ is a family of {downsets}. Let $\Pi_n(\Omega)=\{(A_1,\dots, A_n): {A_i\in\mathcal F,}\ \cup_{i=i}^n A_i=\Omega,~ A_i\cap A_j=\emptyset, ~i\neq j\}$ {denote the class of all measurable partitions of $\Omega$.}
\begin{theorem}\label{Th:sharing} If $\rho_i: \X\to[-\infty,\infty),~i\in[n]$, then 
\begin{align}\label{eq:mg}
\dsquare_{i=1}^n\rho_i(X)=\inf\{x\in\R: \{X>x\}\in \widetilde{\mathcal A}_x\}.
\end{align}
Moreover, if {the right-hand side of \eqref{eq:mg}  admits a minimizer $x^*$}, then the optimal allocation is given by
\begin{equation}\label{Optimal11}
X_i=\left(X-x^*\right)\id_{A_i^*}+y_i^*,~i\in [n],
\end{equation}
where  $(y_1^*, \dots, y_n^*)\in\R^n$ and  $(A_1^*, \dots, A_n^*)\in\Pi_n(\Omega)$ satisfy $\sum_{i=1}^n y_i^*=x^*$ and 
${\{X>x^*\}\cap A_i^*}\in \mathcal A_{y_i^*}^{(i)},~i\in [n]$. Moreover, if $\{\mathcal A_x^{(i)}\}_{x\in\R},~i\in [n]$ are all increasing, then $\{\widetilde{\mathcal A}_x\}_{x\in\R}$ is increasing; if $\{\mathcal A_x^{(i)}\}_{x\in\R},~i\in [n]$ are all decreasing, then $\{\widetilde{\mathcal A}_x\}_{x\in\R}$ is decreasing and left-continuous.
\end{theorem}

Note that the main message in Theorem \ref{Th:sharing} is that the {class of} induced risk measures {is} closed under inf-convolution. {That is, the} inf-convolution {of induced risk measures is the} risk measure induced by the family of {downsets} $\{\widetilde{\mathcal A}_x\}_{x\in\R}$. 
As $\Lambda\VaR^w$ belongs to this class of risk measures, Theorem \ref{Th:sharing} shows that the inf-convolution of $\Lambda\VaR^w$ is an induced risk measure.  Moreover, it also offers the optimal allocation by splitting the tail part of the risk. If $\mathcal A_x^{(i)}=\mathcal A_y^{(i)}$ for all $x,y\in\R,~i\in [n]$, then Theorem \ref{Th:sharing} {reduces} to Theorem 2 of \cite{LMW24}, where  {it was shown} that the inf-convolution of Choquet quantiles is still a Choquet quantile.

Let us now consider  the inf-convolution of $\Lambda\VaR^w$.
We first introduce the following notation.
For $\mathbf \Lambda=(\Lambda_1,\dots,\Lambda_n)$ and $\mathbf w=(w_1,\cdots,w_n)$, define
\begin{align*}
    \mathcal A_{x}^{\mathbf \Lambda, \mathbf w}=\left\{\cup_{i=1}^n A_i: w_i(A_i)\leq \Lambda_i(y_i),~i\in [n],~\text{{where}}~\sum_{i=1}^ny_i=x\right\},~x\in\R.
\end{align*}
Clearly, $\{\mathcal A_{x}^{\mathbf \Lambda, \mathbf w}\}_{x\in\R}$ is a family of downsets.

\begin{proposition}\label{Thmain} For $\Lambda_i\in\mathcal H$ with $\lambda_i^->0$ and continuous capacities $w_1,\dots, w_n$, we have
\begin{align}\label{eq:multiple}\dsquare_{i=1}^n\Lambda_i\VaR^{w_i}(X)=\inf\{x\in\R: \{X>x\}\in \mathcal A_{x}^{\mathbf \Lambda, \mathbf w}\}.
\end{align}
Moreover, if {the right hand of \eqref{eq:multiple}  admits a minimizer  $x^*$}, then the optimal allocation is given by
\begin{equation}\label{Optimal}
X_i=\left(X-x^*\right)\id_{A_i^*}+y_i^*,~i\in [n],
\end{equation}
where  $(y_1^*, \dots, y_n^*)\in\R^n$ and  $(A_1^*, \dots, A_n^*)\in\Pi_n(\Omega)$ satisfy $\sum_{i=1}^n y_i^*=x^*$ and $w_i(\{X>x^*\}\cap A_i^*)\leq \Lambda_i(y_i^*)$.
Additionally,
if $(\Lambda_1,\dots,\Lambda_n)\in\mathcal H_I^n$, then the inf-convolution in \eqref{eq:multiple} is {of the form} $\Lambda\VaR^w$ {for some} $\Lambda\in\mathcal H_I^*$; if $(\Lambda_1,\dots,\Lambda_n)\in\mathcal H_D^n$, then the inf-convolution in \eqref{eq:multiple} is  {of the form}  $\Lambda\VaR^w$ {for some} $\Lambda\in \mathcal H_D^*$.
\end{proposition}

Our result in Proposition \ref{Thmain} shows that the inf-convolution of $\Lambda\VaR^w$ is an induced risk measure as stated in Theorem \ref{Th:sharing}. It further provides a more concrete answer by showing that if all $\Lambda$ functions are monotone in the same direction, then the inf-convolution is still a  $\Lambda\VaR^w$.  But it is not clear whether the inf-convolution of $\Lambda\VaR^w$ for general $\Lambda$ functions is still {of the form} $\Lambda\VaR^w$ or not.

  The allocation \eqref{Optimal} can be understood as follows. {Agent $i$} receives a constant loss $y_i^*$ along with a contingent loss. The contingent component corresponds to the portion of the total risk that exceeds the aggregate cash allocation and is assigned to the agent {on} the set $A_i^*$. The defining {characteristic} of those sets is revealed by the structure of $\mathcal A_{x}^{\mathbf \Lambda, \mathbf w}$: for each agent, {both the capacity of  the set $A_i^*$ on which they suffer the contingent loss and  the total loss exceeding the aggregate cash allocation are restricted by the function $\Lambda_i$. } 

Note that Proposition \ref{Thmain} is  very general, covering the corresponding results in \cite{ELW18}, \cite{ELMW19}, \cite{LMW24}, \cite{LSW24}, and \cite{L24}.  {Moreover,} if all $\Lambda_i$ are constants and $w_i$ are probabilities, then Proposition \ref{Thmain} {reduces} to Theorem 4 of \cite{ELMW19}, where the inf-convolution of $\VaR$ was studied with belief heterogeneity{. If}  all $\Lambda_i$ are constants, then Proposition \ref{Thmain} extends Theorem 2 of \cite{LMW24}, where  the inf-convolution of Choquet quantiles was studied, showing that the inf-convolution of Choquet quantiles is still a Choquet quantile. It provides an answer to {a question posed in Section 3 of}  \cite{LSW24} by showing that the inf-convolution of $\Lambda\VaR^\p$ with belief heterogeneity and monotone $\Lambda$ functions  is {of the form} $\Lambda\VaR^w$ {for  a capacity $w$. This result is stated in the following corollary.}

\begin{corollary}
If  {all functions $\Lambda_i$ are increasing or decreasing and satisfy  $\lambda_i^->0$}, then 
\begin{align*}\dsquare_{i=1}^n\Lambda_i\VaR^{\p_i}(X)=\inf\{x\in\R: \{X>x\}\in \mathcal A_{x}^{\mathbf \Lambda, (\p_1,\dots, \p_n)}\}
\end{align*}
is {of the form  $\Lambda\VaR^w$ for some function $\Lambda$}.
\end{corollary}


\begin{remark}\label{Remark}
   The assumption $\lambda_i^->0$ together with the continuity of $w_i$ is {required so as} to guarantee that $\Lambda_i\VaR^{w_i}(X)<\infty$. Note that $\Lambda_i\VaR^{w_i}(X)<\infty$ holds true if $\Lambda_i\in\mathcal H_I$ and $w_i$ is continuous. Hence, if all $\Lambda_i$ {belong to} $\mathcal H_I$, then the assumption $\lambda_i^->0$ can be removed from Proposition \ref{Thmain}.    
\end{remark}

An immediate consequence of Proposition \ref{Thmain}, Remark \ref{Remark} and Corollary \ref{Cor:motive} is the following result on risk sharing under model uncertainty. 

\begin{corollary}\label{Cor:main} For uncertainty sets $\mathcal P_1$, \dots $\mathcal P_n$ consisting of probability measures and $\Lambda_i\in\mathcal H_I$, if $\overline {\mathbb Q}_i=\sup_{\mathbb Q\in \mathcal P_i}\mathbb Q$ are continuous capacities, then  we have
\begin{align*}\dsquare_{i=1}^n\sup_{\mathbb Q\in \mathcal P_i}\Lambda_i\VaR^{\mathbb Q}(X)=\inf\{x\in\R: \{X>x\}\in \mathcal A_{x}^{\mathbf \Lambda, (\overline {\mathbb Q}_1,\dots, \overline {\mathbb Q}_n)}\}
\end{align*}
is a $\Lambda\VaR^w$ with $\Lambda\in\mathcal H_I^*$,
 and the optimal allocation has form \eqref{Optimal} if the minimizer of the right-hand side of the above equality  exists.  
\end{corollary}
Corollary \ref{Cor:main} shows that the inf-convolution of $\Lambda\VaR$ under ambiguity  with increasing $\Lambda$ is {of the form $\Lambda\VaR^w$ for a capacity} $w$. This result extends the inf-convolution of $\VaR$ under ambiguity in \cite{LMW24} and with heterogeneous beliefs in \cite{ELMW19}. 


Next, let us discuss the finiteness of $\Gamma_{\mathbf \Lambda, \mathbf w}(X):=\inf\{x\in\R: \{X>x\}\in \mathcal A_{x}^{\mathbf \Lambda, \mathbf w}\}$. Let  $\bigwedge_{i=1}^n x_i=\min_{i=1}^n x_i$ and $\bigvee_{i=1}^n x_i=\max_{i=1}^n x_i$. We say a capacity $w$ is \emph{subadditive} if  
$w(A\cup B)\leq w(A)+w(B)$ for all $A,B\in\mathcal F$.  

\begin{proposition}\label{prop:finite2} 
Suppose all $\Lambda_i\in\mathcal H_I$ with $0<\lambda_i^-\leq \lambda_i^+<1$ and all $w_i$ are continuous and subadditive. Then we have the following conclusion.
\begin{enumerate}[(i)]
\item If $\inf_{(A_1,\dots, A_n)\in\Pi_n(\Omega)}\bigwedge_{i=1}^n\left(\frac{w_i(A_i)}{\lambda_i^-}\vee\bigvee_{j\neq i}\frac{w_j(A_j)}{\lambda_j^+}\right)<1$, then $\Gamma_{\mathbf \Lambda, \mathbf w}(X)=-\infty$;
\item If $\inf_{(A_1,\dots, A_n)\in\Pi_n(\Omega)}\bigwedge_{i=1}^n\left(\frac{w_i(A_i)}{\lambda_i^-}\vee\bigvee_{j\neq i}\frac{w_j(A_j)}{\lambda_j^+}\right)>1$, then $\Gamma_{\mathbf \Lambda, \mathbf w}(X)>-\infty$.
\end{enumerate}
\end{proposition}

Note that {each capacity $\overline{\Q}_i=\sup_{\Q\in\mathcal P_i}\Q$ is automatically subadditive but not necessarily continuous}. If $\overline{\Q}_i$ are continuous, then the conclusion in Proposition \ref{prop:finite2} holds for the inf-convolution in Corollary \ref{Cor:main}. Proposition \ref{prop:finite2}  extends some results in  Proposition 3 of \cite{LSW24}, where $w_i=\p_i$ was considered.

\subsection{Risk sharing under ambiguity}

In this subsection, we consider risk sharing {under ambiguity for multiple risk measures of the form $\Lambda\VaR$, where  the uncertainty sets are  induced either by the $\phi$-divergence and the likelihood ratio. By applying} Theorem \ref{cor:upper1}, Proposition \ref{Prop:likelihood1}, Corollary \ref{Cor:motive} and Corollary \ref{Cor:main}, we immediately arrive at the following result.

\begin{proposition}\label{Cor:divergence}
   Suppose $\Lambda_i\in \H_I$ and $\boldsymbol{\mathrm P}=(\p_1,\dots,\p_n)$.
   \begin{enumerate}[(i)]
  \item  If $\phi_i$ is strictly convex, $\lim_{x\to\infty}\frac{\phi_i(x)}{x}=\infty$ and $\delta_i>0$,  then we have 
  \begin{align*}\dsquare_{i=1}^n\sup_{\mathbb Q\in \mathcal P_{\phi_i}(\p_i,\delta_i)}\Lambda_i\VaR^{\mathbb Q}(X)&=\dsquare_{i=1}^n(g_{\phi_i,\delta_i}^{-1}\circ\Lambda_i)\VaR^{\mathbb P_i}(X)\\
  &=\inf\left\{x\in\R: \{X>x\}\in \mathcal A_{x}^{(g_{\phi_1,\delta_1}^{-1}\circ\Lambda_1,\dots, g_{\phi_n,\delta_n}^{-1}\circ\Lambda_n),\boldsymbol{\mathrm P}}\right\},
  \end{align*}
  where {the functions} $g_{\phi_i,\delta_i}$ are given in Proposition \ref{Prop:alpha}. If the minimizer of the right-hand side of the above {identity} exists, then the optimal allocation is given by  \eqref{Optimal}.
   \item  If $0\leq k_{i,1}< 1< k_{i,2}$, then  we have 
  \begin{align*}\dsquare_{i=1}^n\sup_{\mathbb Q\in \mathcal P(\p_i,k_{i,1},k_{i,2})}\Lambda_i\VaR^{\mathbb Q}(X)&=\dsquare_{i=1}^n(g_i^{-1}\circ\Lambda_i)\VaR^{\mathbb P_i}(X)\\
  &=\inf\left\{x\in\R: \{X>x\}\in \mathcal A_{x}^{(g_{1}^{-1}\circ\Lambda_1,\dots, g_{n}^{-1}\circ\Lambda_n),\boldsymbol{\mathrm P}}\right\},
  \end{align*}
  where $g_i(x)=(k_{i,2}x)\wedge (k_{i,1}x+1-k_{i,1})$ for $x\in [0,1]$. If the minimizer of the right-hand side of the above  {identity} exists, then the optimal allocation is given by  \eqref{Optimal}.
  \end{enumerate}
\end{proposition}

{The results in Proposition \ref{Cor:divergence} show that risk sharing with these two uncertainty sets can be transformed into a problem without ambiguity but with heterogeneous beliefs.  The special case $k_{i,1}=0$ in part (ii) of Proposition \ref{Cor:divergence}} coincides with Corollary 2 of \cite{LMWW22} and Proposition 4 of \cite{XH24}. Note that some explicit formulas for the risk sharing with $\Lambda\VaR$ under heterogeneous beliefs were given in Theorems 2-4 of \cite{LSW24}.

{The case $\p_1=\dots=\p_n=\p$ represents the scenario in which all agents agree on the same reference probability but may have} different opinions on the uncertainty sets.  In this case, we can obtain more explicit results using Theorem 2 of \cite{LSW24}.  For $\Lambda_i\in \mathcal H$ and $x\in\R$, let
$$\Lambda^*(x)=1\wedge\sup_{y_1+\dots+y_n=x}\sum_{i=1}^{n}\Lambda_i(y_i).$$
 We say $\Lambda^*$ is attainable if for each $x\in\R$, there exists $(y_1,\dots,y_n)\in\R^n$ satisfying $\sum_{i=1}^n y_i=x$ such that $1\wedge \sum_{i=1}^{n}\Lambda_i(y_i)=\Lambda^*(x)$.
It is shown in the proof of Proposition C.1 in \cite{L24} that if $\Lambda_i\in \mathcal H_I$  is right-continuous and there exist $y_{i,1}$ and $y_{i,2}$ such that $\Lambda_i(y_{i,1})=\lambda_i^-$ and $\Lambda_i(y_{i,2})=\lambda_i^+$, then $\Lambda^*$ is attainable.
Note that $\Lambda^*$ is an increasing function if all $\Lambda_i$ are increasing.
For $x\in\R$, let $$\Lambda_1^*(x)=1\wedge \sup_{y_1+\dots+y_n=x}\sum_{i=1}^{n}g_{\phi_i,\delta_i}^{-1}\circ\Lambda_i(y_i),~\text{and}~\Lambda_2^*(x)=1\wedge \sup_{y_1+\dots+y_n=x}\sum_{i=1}^{n}g_i^{-1}\circ\Lambda_i(y_i).$$
Applying Proposition \ref{Cor:divergence} above and Theorem 2 of \cite{LSW24}, we have the following results. 
   \begin{corollary} Suppose $\Lambda_i\in \H_I$ and $\p$ is atomless.
   \begin{enumerate}[(i)]
  \item  If  If $\phi_i$ is strictly convex, $\lim_{x\to\infty}\frac{\phi_i(x)}{x}=\infty$, $\delta>0$ and $\Lambda_1^*$ is attainable, then
  \begin{align*}\dsquare_{i=1}^n\sup_{\mathbb Q\in \mathcal P_{\phi_i}(\p,\delta_i)}\Lambda_i\VaR^{\mathbb Q}(X)=\Lambda_1^*\VaR^{\mathbb P}(X).
  \end{align*}
   \item  If $0\leq k_{i,1}< 1< k_{i,2}$, and $\Lambda_2^*$ is attainable, then  
  \begin{align*}\dsquare_{i=1}^n\sup_{\mathbb Q\in \mathcal P(\p,k_{i,1},k_{i,2})}\Lambda_i\VaR^{\mathbb Q}(X)=\Lambda_2^*\VaR^{\mathbb P}(X).
  \end{align*}
  \end{enumerate}
   \end{corollary}
   
The above results also imply compact formulas for risk sharing with $\VaR$ under model uncertainty.  For $0<\alpha_i<1$, if $\p$ is atomless, then
\begin{align*}&\dsquare_{i=1}^n\sup_{\mathbb Q\in \mathcal P_{\phi_i}(\p,\delta_i)}\VaR_{\alpha_i}^{\mathbb Q}(X)=\VaR_{\sum_{i=1}^ng_{\phi_i,\delta_i}^{-1}(\alpha_i)}^{\p}(X),\\
&\dsquare_{i=1}^n\sup_{\mathbb Q\in \mathcal P(\p,k_{i,1},k_{i,2})}\VaR_{\alpha_i}^{\mathbb Q}(X)=\VaR_{\sum_{i=1}^ng_i^{-1}(\alpha_i)}^{\p}(X).
\end{align*}

Next, we consider risk sharing with the uncertainty sets $\mathcal P(\p_1,0,Y_1), \dots, \mathcal P(\p_n,0,Y_n)$ defined in \eqref{Eq:likelihood}. Applying  Proposition \ref{Prop:likelihood1}, Corollary \ref{Cor:motive} and Corollary \ref{Cor:main}, we immediately arrive at the following results.
\begin{proposition}\label{Th:likelihood} For  $\Lambda_i\in \H_I$,  we have 
  \begin{align*}\dsquare_{i=1}^n\sup_{\mathbb Q\in \mathcal P(\p_i,0,Y_i)}\Lambda_i\VaR^{\mathbb Q}(X)=\inf\{x\in\R: \{X>x\}\in \mathcal A_{x}^{\mathbf \Lambda, (\overline {\mathbb Q}_1,\dots, \overline {\mathbb Q}_n)}\},
  \end{align*}
  where $\overline{\mathbb Q}_i(A)=1\wedge \mathbb E^{\mathbb P_i}(Y_i\id_{A})$ for all $A\in\mathcal F$.
  Moreover, if the minimizer of right-hand side of the above equality exists, then the optimal allocation is given by \eqref{Optimal}.
\end{proposition} 
  
  Next, we consider {the} special case $\mathcal P_1=\dots=\mathcal P_n$ and offer an explicit expression of the inf-convolution.
  For $x\in\R$, let $$\overline{\Lambda}^*(x)=\sup_{y_1+\dots+y_n=x}\sum_{i=1}^{n}\Lambda_i(y_i).$$ 
   
\begin{corollary}\label{prop:likelihood}  For $\Lambda_i\in \H_I$,  if $\overline{\Lambda}^*$ is attainable and $\p$ is atomless, then 
       \begin{align*}\dsquare_{i=1}^n\sup_{\mathbb Q\in \mathcal P(\p,0,Y)}\Lambda_i\VaR^{\mathbb Q}(X)=\inf\left\{x\in\R: \mathbb E^{\mathbb P}\left(Y\id_{\{X>x\}}\right)\leq \overline{\Lambda}^*(x)\right\}.
  \end{align*}
  Moreover,  if additionally the {right-hand expression has the finite value $x^*$}  and $\overline{\Lambda}^*$ is right-continuous at $x^*$, then the optimal allocation is given by \eqref{Optimal} with  $(y_1^*, \dots, y_n^*)$  satisfying $\sum_{i=1}^{n}\Lambda_i(y_i^*)=\Lambda^*(x^*)$ and $(A_1^*, \dots, A_n^*)$ satisfying $\mathbb E^{\mathbb P}\left(Y\id_{\{X>x^*\}\cap A_i^*}\right)\leq \Lambda_i(y_i^*)$.
\end{corollary}

We notice that if {$\Lambda_i\in \mathcal H_I,~i \in [n]$,} are right-continuous, then $\overline{\Lambda}^*$ is right-continuous; see proposition 8 of \cite{XH24}.

{If we no longer assume that  $\Lambda^*$ is attainable, then we can still obtain the following corollary.
For its statement, we define for  $n\geq 2$,}
$$\Lambda^{\mathbf{y}_{n-1}}(x)=\Lambda_{n}(x-y_{n-1})+
\sum_{i=1}^{n-1}\Lambda_{i}(y_i-y_{i-1}),$$ where   $\mathbf{y}_{n-1}=(y_1,\dots,y_{n-1})$ and $y_0=0$.
\begin{corollary}\label{Prop:111}
 For $\Lambda_i\in \H_I$, if $\p$ is atomless, then we have
       \begin{align*}\dsquare_{i=1}^n\sup_{\mathbb Q\in \mathcal P(\p,0,Y)}\Lambda_i\VaR^{\mathbb Q}(X)=\inf_{\mathbf y_{n-1}\in\R^{n-1}}\inf\left\{x\in\R: \mathbb E^{\mathbb P}\left(Y\id_{{\{X>x\}}}\right)\leq \Lambda^{\mathbf{y}_{n-1}}(x)\right\}.
  \end{align*}    
\end{corollary}

If one of {the functions} $\Lambda_i$ is a constant,  one can easily check that $\overline{\Lambda}^*=\lambda_1^++\dots+\lambda_n^+$ is also a constant.  Using  this fact and the conclusion in Corollary \ref{prop:likelihood}, we obtain the following result.
\begin{corollary}
For $\Lambda_i\in \H_I$, if $\overline{\Lambda}^*$ is attainable, one of $\Lambda_i$ is a constant and $\p$ is atomless, then  we have
       \begin{align*}\dsquare_{i=1}^n\sup_{\mathbb Q\in \mathcal P(\p,0,Y)}\Lambda_i\VaR^{\mathbb Q}(X)=\inf\left\{x\in\R: \mathbb E^{\mathbb P}\left(Y\id_{{\{X>x\}}}\right)\leq \sum_{i=1}^n\lambda_i^+\right\}.
  \end{align*}      
\end{corollary}

\section{Comonotonic risk sharing}\label{Sec:Comonotonic}
In this section, we suppose {that} allocations are \emph{comonotonic}\footnote{For $X_1,\dots,X_n$, we say that they are comonotonic if there exist increasing  functions $f_1,\dots, f_n$  such that $f_i(X_1+\dots+X_n)=X_i$ for all $i \in [n]$.}.
 The set of allocation is 
 \begin{equation*}
\mathbb{A}_n^+(X)=\left\{(X_1,\ldots,X_n)\in \mathbb{A}_n(X): X_1,\dots, X_n~\text{are comonotonic and } \sign(X_i) \text{ are the same} \right\},
\end{equation*}
where $\sign(x)=\id_{\{x\geq 0\}}-\id_{\{x<0\}}$. Note that the restriction that {all $X_i$ have the same sign guarantees} that positive risk $X$ does not generate {a} negative risk allocation $X_i$.
We are interested in the smallest risk exposure and the optimal allocation $(X_1^*,\dots, X_n^*)\in \mathbb{A}_n^+(X)$ for $\rho_1,\dots,\rho_n$ and $X$ as below:
\begin{equation*}
 \boxplus_{i=1}^n \rho_i (X)  = \inf\left\{\sum_{i=1}^n\rho_i(X_i): (X_1,\cdots,X_n)\in \mathbb A_n^+(X)  \right\},
 \end{equation*}
 and 
{$(X_1^*,\dots, X_n^*)$ is a minimizer of the right-hand expression}.

We say {that} a capacity $w$ is \emph{atomless} if there exists a bounded random variable $X$ such that  $w(X>x)$ is continuous over $\R$.

 \begin{theorem}\label{Thmain1}For $\Lambda_i\in\mathcal H_I$  and continuous capacities $w_1,\dots, w_n$, if $\Lambda_i\VaR^{w_i}(X)\geq 0$ for  $i\in [n]$, then we have
\begin{align}\label{eq:multiple11}\boxplus_{i=1}^n\Lambda_i\VaR^{w_i}(X)=\min_{i \in [n]}\Lambda_i\VaR^{w_i}(X),
\end{align}
and {one optimal allocation is of the form} $X_{i_0}=X$ and $X_i=0$ for $i\neq i_0$, where $$i_0\in \argmin_{i \in [n]}\Lambda_i\VaR^{w_i}(X).$$
Additionally, if {all $\Lambda_i$ are constant on} $(-\infty,0)$,   then \eqref{eq:multiple11} holds true for all $X\in\X$.  Moreover, if  {$w_1=\dots=w_n$, $w_1$ is} atomless and $\Lambda_1=\dots=\Lambda_n$, then the condition {that all $\Lambda_i$ are  constant on $(-\infty,0)$ is both  necessary and sufficient} for the validity of \eqref{eq:multiple11} for all $X\in\X$. 
\end{theorem}

The conclusion in Theorem \ref{Thmain1} extends the results in Theorem 11 of \cite{XH24}, {where the comonotonic risk sharing problem  without model ambiguity and with homogeneous beliefs, i.e., the case $w_1=\dots=w_n=\p$, was studied}. Moreover,  the {final part of Theorem \ref{Thmain1} implies that \eqref{eq:multiple11} may no longer hold if $\Lambda_i\VaR^{w_i}(X)<0$ for some $i \in [n]$, because $\Lambda_i\VaR^{w_i}(X)<0$ cannot guarantee that $\Lambda_i$ is a constant on $(-\infty,0)$. Next, our}   result suggests that the optimal comonotonic allocation is allocating all risk to the agent with the smallest $\Lambda\VaR$, and allocating nothing to {all the} others. This is consistent with the optimal allocation for $\VaR$ with ambiguity sets characterized by Wasserstein {balls}; see Theorem 6 of \cite{LMWW22}. 

If all {$\Lambda_i$ are constant}, then \eqref{eq:multiple11} holds true for all $X\in\mathcal X$. This is because $\Lambda\VaR^{w_i}$ {then reduces to a Choquet quantile}, and Choquet quantiles {are known to} satisfy ordinality: $\VaR_{\alpha}^{w_i}(f(X))=f(\VaR_{\alpha}^{w_i}(X))$ for all increasing and continuous functions $f$; see \cite{C07} and \cite{LMW24} for more details on ordinality and risk sharing with Choquet quantiles.   

An immediate consequence of Theorem \ref{Thmain1} is the {following proposition on comonotonic risk sharing with ambiguity sets}.

 \begin{proposition}\label{Prop:comonotonic}
 For $\Lambda_i\in\mathcal H_I$  and uncertainty sets $\mathcal P_1, \dots, \mathcal P_n$, if $\overline{\Q}_i=\sup_{\Q\in\mathcal P_i}\Q$ are continuous and  $\Lambda_i\VaR^{\overline{\Q}_i}(X)\geq 0$ for all $i \in [n]$, then we have
\begin{align*}\boxplus_{i=1}^n\sup_{\Q\in\mathcal P_i}\Lambda_i\VaR^{\Q}(X)=\min_{i \in [n]}\Lambda_i\VaR^{\overline{\Q}_i}(X),
\end{align*}
and {one optimal allocation is of the form} $X_{i_0}=X$ and $X_i=0$ for $i\neq i_0$, where $$i_0{\in}\argmin_{i \in [n]}\Lambda_i\VaR^{\overline{\Q}_i}(X).$$
\end{proposition}

Applying Proposition \ref{Prop:comonotonic}, we  obtain more explicit formulas for some specific uncertainty sets in the following corollary.
\begin{corollary} Let $\Lambda_i\in\mathcal H_I$ and suppose $\Lambda_i\VaR^{\overline{\Q}_i}(X)\geq 0$ for all $i \in [n]$.
\begin{enumerate}[(i)]
\item If $\mathcal P_i=\mathcal P_{\phi_i}(\p_i,\delta_i)$, then 
    \begin{align*}\boxplus_{i=1}^n\sup_{\Q\in\mathcal P_i}\Lambda_i\VaR^{\Q}(X)=\min_{i \in [n]}(g_{\phi_i,\delta_i}^{-1}\circ\Lambda_i)\VaR^{\p_i}(X).
\end{align*}
\item If $\mathcal P_i=\mathcal P(\p_i,k_{i,1},k_{i,2})$ with $0\leq k_{i,1}<1<k_{i,2}$, then 
    \begin{align*}\boxplus_{i=1}^n\sup_{\Q\in\mathcal P_i}\Lambda_i\VaR^{\Q}(X)=\min_{i \in [n]}(g_{i}^{-1}\circ\Lambda_i)\VaR^{\p_i}(X),
\end{align*}
where $g_i(x)=(k_{i,2}x)\wedge (k_{i,1}x+1-k_{i,1}),~x\in [0,1]$.
\item If $\mathcal P_i=\mathcal P(\p_i,0,Y_i)$ with $Y_i\geq 0$ and $\mathbb E^\p(Y_i)>1$, then 
    \begin{align*}\boxplus_{i=1}^n\sup_{\Q\in\mathcal P_i}\Lambda_i\VaR^{\Q}(X)=\min_{i \in [n]}\Lambda_i\VaR^{\overline{\Q}_i}(X),
\end{align*}
where $\overline{\Q}_i(A)=1\wedge\mathbb E^{\p_i}(Y\id_A)$ for all $A\in\mathcal F$.
\end{enumerate}
\end{corollary}
\section{Conclusion}

{This paper investigates robust models of $\Lambda\VaR$ where the ambiguity sets are composed of probability measures. We demonstrate that when $\Lambda$ is increasing, the robust $\Lambda\VaR$ corresponds to a $\Lambda\VaR$ defined under a capacity, which generally does not constitute a probability measure. We establish several key properties of this extended risk measure, including monotonicity, cash-subadditivity, and quasi-star-shapedness. Furthermore, we provide an equivalent representation of $\Lambda\VaR^w$ for monotone $\Lambda$ using families of downsets.}

{Our analysis yields broad results in the context of risk sharing, showing that risk measures induced by families of downsets are closed under inf-convolution. This implies that the inf-convolution of $\Lambda\VaR^w$ results in another risk measure of the same type. Specifically, we prove that the class of risk measures  of the form $\Lambda\VaR^w$ with a monotone function  $\Lambda$ is stable under inf-convolution. However, whether this property holds for non-monotone $\Lambda$ remains an open question and warrants further investigation.}

{Additionally, we derive risk-sharing results under uncertainty sets defined by $\phi$-divergence and likelihood ratios.}




\appendix

\section{Proof of Section \ref{sec:properties}}\label{Appendix:Sec3}
In this section, we give  all of the proofs of the results in Section \ref{sec:properties}.

{\bf Proof of Proposition \ref{Prop:properties}}. The claims in (i) and (ii) follow directly from the definition. Next, we focus on (iii). To show the quasi-star-shapedness of $\Lambda\VaR^w$, we need to prove that for $X\in\X$ and $t\in\R$, $\Lambda\VaR^{w}(\lambda X+(1-\lambda)t)\leq \max(\Lambda\VaR^{w}(X),t)$ holds for $\lambda\in [0,1]$. Clearly, it holds for $\lambda=0, 1$.  Next, we suppose $\lambda\in (0,1)$. If $\Lambda\VaR^{w}(X)=\infty$, then it holds obviously. Next, we assume $\Lambda\VaR^{w}(X)\in [-\infty,\infty)$.  For $y>\max(\Lambda\VaR^{w}(X),t)$, it follows from the definition of $\Lambda\VaR^{w}$ and the monotonicity of $\Lambda$ that $w(X>y)\leq \Lambda(y)$, which implies
\begin{align*}
    w(\lambda X+(1-\lambda)t>y)=w(X>y+(1-\lambda)(y-t)/\lambda)\leq w(X>y)\leq \Lambda(y).
\end{align*}
Consequently, $\Lambda\VaR^{w}(\lambda X+(1-\lambda)t)\leq y$.  Using the arbitrariness of $y$, we have $\Lambda\VaR^{w}(\lambda X+(1-\lambda)t)\leq \max(\Lambda\VaR^{w}(X),t)$.  The quasi-star-shapedness for $\Lambda\VaR^{+,w}$ can be shown similarly. The \trd{details} are omitted.

Finally, we show (iv).  By definition, $\Lambda\VaR^{w}(X+m)=\Lambda^{m}\VaR^{w}(X)+m$, where $\Lambda^m(x)=\Lambda(x+m)$. For $m\geq 0$, if $\Lambda$ is increasing, then $\Lambda^m\geq \Lambda$. Using the conclusion in (ii), we have $\Lambda\VaR^{w}(X+m)\leq \Lambda\VaR^{w}(X)+m$, showing cash-subadditivity. If $\Lambda$ is decreasing, then $\Lambda^m\leq \Lambda$. Using the conclusion in (ii), we have $\Lambda\VaR^{w}(X+m)\geq \Lambda\VaR^{w}(X)+m$, showing cash-supadditivity.  The proof for $\Lambda\VaR^{+,w}$ is similar and it is omitted.  \qed



{\bf  Proof of Proposition \ref{prop:motive}}. We show first that $\Lambda\VaR^{w}(X)\le \Lambda\VaR^{\overline w}(X)$ for all $w\in W$. Indeed, since $\{x\in \R: \overline{w}(X>x)\leq \Lambda(x)\}\subseteq \{x\in \R: w(X>x)\leq \Lambda(x)\}$, 
\begin{align*}
    \Lambda\VaR^{w}(X)=\inf \{x\in \R: w(X>x)\leq \Lambda(x)\}\le \inf \{x\in \R: \overline{w}(X>x)\leq \Lambda(x)\}=\Lambda\VaR^{\overline w}(X).
\end{align*}
To show that $\sup_{w\in W}\Lambda\VaR^w$ dominates $\Lambda\VaR^{\overline{w}}$, we take $y<\Lambda\VaR^{\overline w}(X)$. Then $\overline{w}(X>y)>\Lambda(y)$, and so there exists $w_y\in W$ such that 
$$w_y(X>y)>\Lambda(y)\ge\Lambda(x)\qquad\text{for all $x\le y$,}$$
where in the final step we have used the assumption that $\Lambda$ is increasing. It follows that $\Lambda\VaR^{w_y}(X)\ge y$. Taking the limit $y\uparrow \Lambda\VaR^{\overline w}(X)$ now yields 
$$\sup_{w\in W}\Lambda\VaR^w(X)\ge\Lambda\VaR^{\overline{w}}(X).$$
The proof of the inequalities for $\Lambda\VaR^+$ is analogous.\qed

    {\bf Proof of Proposition \ref{prop:CQ}}. 
   Let $x_0=\Lambda\VaR^{w}(X)$. First, suppose $x_0\in \R$. By the definition of $\Lambda\VaR^w$ and the monotonicity of $\Lambda$, we have  $w(X>x)\leq \Lambda(x)$ if $x>x_0$ and $w(X>x)>\Lambda(x)$ if $x<x_0$. This implies $\VaR_{\Lambda(x)}^{w} (X)\geq x$ if $x<x_0$ and $\VaR_{\Lambda(x)}^{w} (X)\leq x$ if $x>x_0$. Hence,  $\VaR_{\Lambda(x)}^{w} (X)\vee x=\VaR_{\Lambda(x)}^{w} (X)\geq x_0$ if $x<x_0$ and $\VaR_{\Lambda(x)}^{w} (X)\vee x=x$ if $x>x_0$. Consequently,
   $$\Lambda\VaR^{w}(X)=\inf_{x\in\R} \left\{\VaR_{\Lambda(x)}^{w} (X) \vee x\right\}.$$
   Moreover, the fact that $\VaR_{\Lambda(x)}^{w} (X)\geq x$ if $x<x_0$ and $\VaR_{\Lambda(x)}^{w} (X)\leq x$ if $x>x_0$ also implies 
   $\VaR_{\Lambda(x)}^{w} (X)\wedge x=x$ if $x<x_0$ and  $\VaR_{\Lambda(x)}^{w} (X)\wedge x=\VaR_{\Lambda(x)}^{w} (X)\leq x_0$ if $x>x_0$. Consequently,  $$\Lambda\VaR^{w}(X)=\sup_{x\in \mathbb{R}} \left\{\VaR_{\Lambda(x)}^{w} (X) \wedge x\right\}.$$

   Next, we consider the case $x_0=-\infty$. By the definition of $\Lambda\VaR^w$ and the monotonicity of $\Lambda$, we have  $w(X>x)\leq \Lambda(x)\leq \Lambda(y)$ for all $x\leq y$ with $y\in\R$, which implies $\VaR_{\Lambda(y)}^{w} (X)=-\infty$ for all $y\in\R$. Hence, the two representations hold.

   Finally, we focus on the case $x_0=\infty$. Using the definition of $\Lambda\VaR^w$ and the monotonicity of $\Lambda$, we have 
   $w(X>x)>\Lambda(x)$ for all $x\in\R$. Hence, $\VaR_{\Lambda(x)}^{w} (X)=\infty$ for all $x\in\R$ and the two representations hold. 

    The representations for $\Lambda\VaR^{+,w}(X)$ can be shown similarly. Hence, it is omitted. \qed

{\bf Proof of Theorem \ref{Pro:representation}}. 
Case (i). For  $\Lambda\in\mathcal H_I^*$ and a capacity $w$, let $\mathcal A_x=\{A\in\mathcal F: w(A)\leq \Lambda(x)\}$. Then $\{\mathcal A_x\}_{x\in\R}$ is an increasing  family of downsets and we have 
$$\Lambda\VaR^w(X)=\inf\{x\in\R:\{X>x\}\in \mathcal A_x\},~X\in\X.$$
This shows the ``only if'' part.

Next, we focus on the ``if'' part. Suppose there exists an increasing family of downsets $\{\mathcal A_x\}_{x\in\R}$ and  $\mathcal A_x\subsetneq\mathcal F$ for all $x\in\R$.  If $\mathcal A_x=\mathcal A_y$ for all $x,y\in\R$, then let $\Lambda=\alpha\in (0,1)$, and $w(A)=0$ for $A\in \mathcal A_0$  and $w(A)=1$ for $A\in \mathcal F\setminus\mathcal A_0$. It follows that
$$\Lambda\VaR^w(X)=\inf\{x\in\R:\{X>x\}\in \mathcal A_x\},~X\in\X.$$

Next, we consider the case $\mathcal A_x\neq \mathcal A_y$ for some $x,y\in\R$.
Let $\{(a_i,b_i)\}_{i\in I}$ be the collection of the largest intervals such that \trd{$a_i<b_i$ and $\mathcal A_x=\mathcal A_y$ for all $x,y\in (a_i,b_i)$}. Note that these intervals are disjoint. Let $I_1=\{i\in I: \mathcal A_{a_i}=\mathcal A_{c_i}, a_i>-\infty\}$ with $c_i=(a_i+1)\wedge ((a_i+b_i)/2)$.  Note that if $b_i=\infty$, then  $c_i\in \R$. Let $I_2=\{i\in I: a_i=-\infty\}$. Let $f(x)=\frac{1}{2}+\frac{1}{\pi}\arctan(x),~x\in\R$.
Then for $x\in\R$, define 
\begin{align}\label{Eq:ReLambda}\Lambda(x)=f(x)\id_{\R\setminus \mathbb B}(x)+\sum_{i\in I_1} f(a_i)\id_{[a_i,b_i)}(x)+\sum_{i\in I\setminus (I_1\cup I_2)} f(c_i)\id_{(a_i,b_i)}(x)+\sum_{i\in I_2}f(b_i-1)\id_{(-\infty,b_i)}(x),
\end{align}
where $\mathbb B=\left(\cup_{i\in I_1}[a_i,b_i)\right)\bigcup \left(\cup_{i\in I\setminus I_1}(a_i,b_i)\right)$.

 Next, we construct $w$. For $x\in\R$, let $$\mathcal A_x^+:=\bigcap_{y>x}\mathcal A_y, ~\mathcal A_x^-:=\bigcup_{y<x}\mathcal A_y.$$ 
 Define $w:\mathcal F\to [0,1]$ as \trd{follows}. Let
 $w(A)=\Lambda(x)$ if $A\in \mathcal A_x\setminus \mathcal A_x^-$ for $x\in\R$; $w(A)=\frac{\Lambda(x)+\Lambda(x+)}{2}$ if $A\in \mathcal A_x^+\setminus \mathcal A_x$ for $x\in\R$; $w(A)=0$ if $A\in \bigcap_{x\in\R} \mathcal A_x$; $w(A)=1$ if $A\in \mathcal F \setminus\left(\bigcup_{x\in\R} \mathcal A_x\right)$. 
 
 Clearly, $w$ is well-defined. Under this definition, $w(\emptyset)=0$. \trd{By our assumption $\mathcal A_x\subsetneq\mathcal F$ for all $x\in\R$, we have $w(\Omega)=1$}. Suppose $A\subseteq B$. If $B\in \mathcal F \setminus\left(\bigcup_{x\in\R} \mathcal A_x\right)$, then $w(B)=1$, implying $w(A)\leq w(B)$. If $B\in \bigcap_{x\in\R} \mathcal A_x$, then \trd{using the fact that each $\mathcal A_x$ is a downset}, we have $A\in \bigcap_{x\in\R} \mathcal A_x$. Hence, $w(A)=w(B)=0$. If  $B\in \mathcal A_x\setminus \mathcal A_x^-$ for some $x\in\R$, using the fact that $\mathcal A_x$ is a downset, we have  $A\in \mathcal A_x$. If $A\in \mathcal A_x\setminus \mathcal A_x^-$, we have $w(A)=w(B)=\Lambda(x)$. Or else, there exists $x_1<x$ such that $A\in  \mathcal A_{x_1}\setminus \mathcal A_{x_1}^-$, $A\in  \mathcal A_{x_1}^+\setminus \mathcal A_{x_1}$ or $A\in\bigcap_{y\in\R} \mathcal A_y$. In any of those three cases, $w(A)\leq w(B)$. If  $B\in \mathcal A_{x}^+\setminus \mathcal A_{x}$, we have $A\in \mathcal A_{x}^+$. Using the same argument as above,  we can show that $w(A)\leq w(B)$.
 Hence, $w$ is a capacity. 
 
 For  $x\in\R$, let $x^*=\sup\{y\in\R: \Lambda(y)=\Lambda(x)\}$. If $x^*=\infty$, then $\Lambda(y)=\Lambda(x)$ for all $y\geq x$, implying $\mathcal A_y=\mathcal A_x$ for all $y\geq x$. Hence,   $$\{A\in\mathcal F: w(A)\leq \Lambda(x)\}=\bigcup_{y\geq x} \mathcal A_y=\mathcal A_x.$$ 
 If $x^*=x$, then $\Lambda(x)=\frac{\Lambda(x)+\Lambda(x+)}{2}$ whenever $\Lambda(x)=\Lambda(x+)$; or else, $\Lambda(x)<\frac{\Lambda(x)+\Lambda(x+)}{2}$. Hence, we have
 $$\mathcal A_x\subseteq \{A\in\mathcal F: w(A)\leq \Lambda(x)\}\subseteq \mathcal A_x^+.$$ 
 If $x^*\in (x,\infty)$, then there exists some $i\in I$ such that $x^*=b_i$ and $x\in [a_i,b_i)$. By the expression of $\Lambda$, we have $\Lambda(x^*)>\Lambda(x)$. Hence,
  $$\{A\in\mathcal F: w(A)\leq \Lambda(x)\}=\bigcup_{y\in [x,x^*)} \mathcal A_y=\mathcal A_x.$$ 
In summary, for all $x\in\R$, we have
    $$\mathcal A_x\subseteq \{A\in\mathcal F: w(A)\leq \Lambda(x)\}\subseteq \mathcal A_x^+.$$ 
 Consequently, using our constructed $\Lambda$ and $w$, we have for $X\in\X$,
 \begin{align*}\inf\{x\in\R:\{X>x\}\in \mathcal A_x^+\}\leq \inf\{x\in\R:w(X>x)\leq \Lambda(x)\}\leq \inf\{x\in\R:\{X>x\}\in \mathcal A_x\}.
\end{align*}
Let $x^*:=\inf\{x\in\R:\{X>x\}\in \mathcal A_x\}$.
If $x^*=-\infty$, then $\inf\{x\in\R:\{X>x\}\in \mathcal A_x^+\}=-\infty$.
 If $x^*\in\R$, by definition, we have $\{X>x\}\notin \mathcal A_x$ for all $x<x^*$. Suppose  there exists $x_0\in\R$, such that $x_0<x^*$ and $\{X>x_0\}\in \mathcal A_{x_0}^+$. Then $\{X>x_0\}\in \mathcal A_y$ for all $y>x_0$. Note that $\mathcal A_y$ is a downset and $\{X>y\}\subseteq \{X>x_0\}$ for  $y>x_0$. Hence, we have $\{X>y\}\in \mathcal A_y$ for  $y\in (x_0,x^*)$, leading to a contraction. Hence, $\{X>x\}\notin \mathcal A_x^+$ for all $x<x^*$, implying $\inf\{x\in\R:\{X>x\}\in \mathcal A_x^+\}\geq x^*$.  Hence, $\inf\{x\in\R:\{X>x\}\in \mathcal A_x^+\}=x^*$.
 If  $x^*=\infty$, then $\{X>x\}\notin \mathcal A_x$ for all $x\in\R$. Using the same argument for the case $x^*\in\R$, we can show that
$\inf\{x\in\R:\{X>x\}\in \mathcal A_x^+\}=\infty$. Hence, we have, for $X\in\X$,
$$\inf\{x\in\R:\{X>x\}\in \mathcal A_x\}=\inf\{x\in\R:\{X>x\}\in \mathcal A_x^+\},$$
which implies
\begin{align}\label{Eq:conclusion}\Lambda\VaR^{w}(X)=\inf\{x\in\R:\{X>x\}\in \mathcal A_x\}.
\end{align}

For the case of $\mathcal A_x=\mathcal F$ for some $x\in\R$, if $\bigcap_{x\in\R} \mathcal A_x=\mathcal F$, then let $\Lambda=1$. We can check that \eqref{Eq:conclusion} holds for any capacity $w$. 

Otherwise,  there exists $x_0\in\R$ such that $\mathcal A_{x}=\mathcal F$ for $x>x_0$ and $\mathcal A_x\subsetneq \mathcal F$ for $x<x_0$. Define $\Lambda(x)=1$ for $x>x_0$.  Then we construct $\Lambda(x), x\leq x_0$, and $w$  the same as above. Using the same argument, we can show that \eqref{Eq:conclusion} holds. 

Case (ii). For a left-continuous  $\Lambda\in\mathcal H_D^*$ and a capacity $w$, let $\mathcal A_x=\{A\in\mathcal F: w(A)\leq \Lambda(x)\}$. Then $\{\mathcal A_x\}_{x\in\R}$ is a decreasing  and left-continuous family of downsets and we have 
$$\Lambda\VaR^w(X)=\inf\{x\in\R:\{X>x\}\in \mathcal A_x\},~X\in\X.$$
This shows the ``only if'' part.

Next, we focus on the ``if'' part.  Suppose there exists a decreasing and left-continuous family of downsets $\{\mathcal A_x\}_{x\in\R}$ and  $\mathcal A_x\subsetneq\mathcal F$ for all $x\in\R$. If $\mathcal A_x=\mathcal A_y$ for all $x,y\in\R$, then let $\Lambda=\alpha\in (0,1)$, and $w(A)=0$ for $A\in \mathcal A_0$  and $w(A)=1$ for $A\in \mathcal F\setminus\mathcal A_0$. Then the claim holds.

Next, we suppose $\mathcal A_x\neq\mathcal A_y$ for some $x,y\in\R$.  Let $\{(a_i,b_i)\}_{i\in I}$ be the collection of the largest intervals such that \trd{$a_i<b_i$ and $\mathcal A_x=\mathcal A_y$ for all $x,y\in (a_i,b_i)$}. Let $I_1'=\{i\in I: b_i=\infty\}$.  For $f(x)=\frac{1}{2}+\frac{1}{\pi}\arctan(-x),~x\in\R$, define
\begin{align*}\Lambda(x)=f(x)\id_{\R\setminus (\cup_{i\in I}(a_i,b_i])}(x)+\sum_{i\in I\setminus I_1'} f(b_i)\id_{(a_i,b_i]}(x)+\sum_{i\in I_1'}f(a_i+1)\id_{(a_i,\infty)}(x).
\end{align*}
Clearly, $\Lambda\in \mathcal H_D^*$ and it is left-continuous.
For $x\in\R$, let $$\hat{\mathcal A}_x^+:=\bigcup_{y>x}\mathcal A_y.$$ 
Define $w:\mathcal F\to [0,1]$ as $w(A)=\Lambda(x)$ if $A\in \mathcal A_x\setminus \hat{\mathcal A}_x^+$ for $x\in\R$;  $w(A)=0$ if $A\in \bigcap_{x\in\R} \mathcal A_x$; $w(A)=1$ if $A\in \mathcal F \setminus\left(\bigcup_{x\in\R} \mathcal A_x\right)$. As $\{\mathcal A_x\}_{x\in\R}$ is left-continuous, we have $\left(\bigcap_{y<x} A_y\right)\setminus \mathcal A_x=\emptyset$ for $x\in\R$. Hence, $w$ is well-defined. Using similar argument as case (i), we can show that $w$ is a capacity.

For  $x\in\R$, let $x^*=\inf\{y\in\R: \Lambda(y)=\Lambda(x)\}$. If $x^*=-\infty$, then $\Lambda(y)=\Lambda(x)$ for all $y\leq x$, implying $\mathcal A_y=\mathcal A_x$ for all $y\leq x$. Hence,   $\{A\in\mathcal F: w(A)\leq \Lambda(x)\}=\mathcal A_x.$
 If $x^*=x$, then $\{A\in\mathcal F: w(A)\leq \Lambda(x)\}=\mathcal A_x.$
 If $x^*\in (-\infty, x)$, then there exists some $i\in I$ such that $x^*=a_i$ and $x\in (a_i,b_i]$. By the construction of $\Lambda$, we have  $\Lambda(x^*)>\Lambda(x)$. Hence,
  $$\{A\in\mathcal F: w(A)\leq \Lambda(x)\}=\bigcup_{y\in (x^*,x]} \mathcal A_y=\mathcal A_x.$$ 
We conclude that for all $x\in\R$,
    $\{A\in\mathcal F: w(A)\leq \Lambda(x)\}=\mathcal A_x,$
    implying $$\Lambda\VaR^w(X)=\inf\{x\in\R:\{X>x\}\in \mathcal A_x\},~X\in\X.$$
 This completes the proof for the case $\mathcal A_x\subsetneq\mathcal F$ for all $x\in\R$. 

If there exists $x_0\in\R$ such that $\mathcal A_{x_0}=\mathcal F$, then by choosing \trd{$\Lambda=1$} and  a capacity $w$, we have
$$\Lambda\VaR^w(X)=\inf\{x\in\R:\{X>x\}\in \mathcal A_x\}=-\infty,~X\in\X.$$
This completes the proof.
\qed

{\bf Proof of Proposition \ref{Prop:downsets}}. Using the notation in the proof of Theorem \ref{Pro:representation}, we only need to show 
$$\inf\{x\in\R:\{X>x\}\in \mathcal A_x\}=\inf\{x\in\R:\{X>x\}\in \mathcal A_x^+\}=\inf\{x\in\R:\{X>x\}\in \mathcal A_x^-\}.$$
Note that the first equality has been proved in the proof of Theorem \ref{Pro:representation}. We next show 
$$\inf\{x\in\R:\{X>x\}\in \mathcal A_x\}=\inf\{x\in\R:\{X>x\}\in \mathcal A_x^-\}.$$
Clearly,  $x^*:=\inf\{x\in\R:\{X>x\}\in \mathcal A_x\}\leq \inf\{x\in\R:\{X>x\}\in \mathcal A_x^-\}:=y^*.$
If $x^*=\infty$, then $y^*=x^*=\infty$. If $x^*\in\R$, then we have $\{X>x\}\in\mathcal A_x$ for $x>x^*$. Suppose $y^*>x^*$. Then we have
$\{X>x\}\notin\mathcal A_x^-$ for $x^*<x<y^*$. We choose  $x_1,x_2\in\R$ such that $x^*<x_1<x_2<y^*$. Note that $\mathcal A_{x_1}$ is a downset.  Using the fact that $\{X>x_1\}\in\mathcal A_{x_1}$ and $\{X>x_2\}\subseteq \{X>x_1\}$, we have $\{X>x_2\}\in\mathcal A_{x_1}$. By $\mathcal A_{x_1}\subseteq \mathcal A_{x_2}^-$, we further have $\{X>x_2\}\in\mathcal A_{x_2}^-$, which contradicts the fact $\{X>x\}\notin\mathcal A_x^-$ for $x^*<x<y^*$. Hence, $y^*=x^*$. The case $x^*=-\infty$ can be shown similarly. This completes the proof.  \qed
\section{Proof of Section \ref{Sec:robustVaR}.}

This section is devoted to the proofs of  all the results in Section \ref{Sec:robustVaR}.

{\bf Proof of Proposition \ref{Prop:alpha}}. Clearly, for any $A\in\mathcal F$, we have $\overline{\mathbb Q}(A)=0$ if $\p(A)=0$, and $\overline{\mathbb Q}(A)=1$ if $\p(A)=1$. Next, we suppose $0<\p(A)<1$. Let $\mathcal F_A=\{\emptyset, A,A^c,\Omega\}$. For $\Q\in \mathcal P_\phi(\p,\delta)$, define  $\Q_0$ by $\frac{d\Q_0}{\d \p}=\E^\p\left(\frac{\d \Q}{\d\p}\mid \mathcal F_A\right)$. 
Using  Jensen's inequality, we have 
$$\phi\left(\E^\p\left(\frac{\d \Q}{\d\p}\mid \mathcal F_A\right)\right)\leq \E^\p\left(\phi\left(\frac{\d \Q}{\d\p}\right)\mid\mathcal F_A\right),$$
which further implies 
$$\mathbb E^{\p}\left(\phi\left(\E^\p\left(\frac{\d \Q}{\d\p}\mid \mathcal F_A\right)\right)\right)\leq \E^\p\left(\phi\left(\frac{\d \Q}{\d\p}\right)\right)\leq \delta.$$
Hence, $\Q_0\in \mathcal P_\phi(\p,\delta)$ and $\Q_0(A)=\Q(A)$. Note that  \trd{$\frac{d\Q_0}{\d \p}=x\id_A+y\id_{A^c}$, where $x,y\geq 0$,}
$x\p(A)+y\p(A^c)=1$ and $\phi(x)\p(A)+\phi(y)\p(A^c)\leq \delta$. Hence,
\begin{align*}\overline{\mathbb Q}(A)&=\sup_{\{(x,y): x,y\geq 0, x\p(A)+y\p(A^c)=1, \phi(x)\p(A)+\phi(y)\p(A^c)\leq \delta\}}x\p(A)\\
&=\sup_{\{1\leq x\leq 1/\p(A): \phi(x)\p(A)+\phi(\frac{1-x\p(A)}{1-\p(A)})(1-\p(A))\leq \delta\}}x\p(A)\\
&=\sup\{t\in [\p(A),1]: \p(A)\phi(t/\p(A))+(1-\p(A))\phi((1-t)/(1-\p(A)))\leq \delta\}\\
&=g_{\phi,\delta}(\p(A)).
\end{align*}

 Next, we focus on the properties of $g_{\phi,\delta}$. Let $f(x,t):=x\phi(t/x)+(1-x)\phi((1-t)/(1-x)), (x,t)\in(0,1)\times [x,1]$. If $\phi$ is strictly convex, then for fixed $x\in (0,1)$, $f(x,t)$ is continuous and strictly increasing \trd{in $t\in [x,1]$}.  The strict convexity of $\phi$ also implies that $f(x,1)=x\phi(1/x)+(1-x)\phi(0)$ is strictly decreasing and continuous \trd{in} $x\in (0,1)$. Note that  $\lim_{x\uparrow 1}f(x,1)=\phi(1)<\delta$ and $\lim_{x\downarrow 0} f(x,1)=\lim_{x\downarrow 0}(x\phi(1/x)+(1-x)\phi(0))=\infty$. Hence, there exists unique $x_{\delta}$ such that $f(x_{\delta},1)=\delta$, which implies $f(x,1)\leq\delta$ for $x\geq x_\delta$ and $f(x,1)>\delta$ for $0<x<x_\delta$.  For $0<x<x_\delta$, using the fact that $f(x,x)=x\phi(1)=0<\delta$, there exists unique $t\in (x,1)$ such that $f(x,t)=\delta$. Consequently, $g_{\phi,\delta}(x)=1$ for $x\in [x_{\delta},1]$ and $g_{\phi,\delta}(x)\in (x,1)$ for $x\in (0,x_\delta)$.

 For $0<x_1<x_2\leq x_{\delta}$, it follows from the strict convexity of $\phi$ that $f(x_1,t)>f(x_2,t)$ for $t\in [x_2,1]$, which implies $f(x_1,g_{\phi,\delta}(x_2))>f(x_2,g_{\phi,\delta}(x_2))=\delta$. Hence, $g_{\phi,\delta}(x_1)<g_{\phi,\delta}(x_2)$, showing the monotonicity of $g_{\phi,\delta}$.

 Now suppose $0<x_1<x_2<x_{\delta}$. For $0<\epsilon<1-g_{\phi,\delta}(x_2)$, we have
$f(x_1,g_{\phi,\delta}(x_1)+\epsilon)>\delta$. 
Note that $\lim_{x_2\downarrow x_1 }\sup_{t\in [x_2,1]}\mid f(x_2,t)-f(x_1,t)\mid =0$. Hence, there exists $\eta>0$ such that for $x_1<x_2<x_1+\eta$, $f(x_2,g_{\phi,\delta}(x_1)+\epsilon)>\delta$, implying $g_{\phi,\delta}(x_1)<g_{\phi,\delta}(x_2)<g_{\phi,\delta}(x_1)+\epsilon$.
Hence, $\lim_{x_2\downarrow x_1}g_{\phi,\delta}(x_2)=g_{\phi,\delta}(x_1)$.

For $0<x_1<x_2\leq x_{\delta}$ and $0<\epsilon<(g_{\phi,\delta}(x_2)-x_2)/2$, we have $f(x_2,g_{\phi,\delta}(x_2)-\epsilon)<\delta$. Using the fact that $\lim_{x_1\uparrow x_2}\sup_{t\in [x_2,1]}\mid f(x_2,t)-f(x_1,t)\mid =0$, there exists $\eta>0$ such that $f(x_1,g_{\phi,\delta}(x_2)-\epsilon)<\delta$ for $x_2-\eta<x_1<x_2\leq x_{\delta}$. Hence, $g_{\phi,\delta}(x_2)-\epsilon<g_{\phi,\delta}(x_1)<g_{\phi,\delta}(x_2)$. Letting $\epsilon\to 0$, we have 
$\lim_{x_1\uparrow x_2}g_{\phi,\delta}(x_1)=g_{\phi,\delta}(x_2)$.  

Finally, we consider the continuity of $g_{\phi,\delta}$ at $x=0$.
 For any $\epsilon\in (0,x_\delta)$, $\lim_{x\downarrow 0} f(x,\epsilon)=\lim_{x\downarrow 0}(x\phi(\epsilon/x)+(1-x)\phi((1-\epsilon)/(1-x)))=\infty.$
 Hence, there exists $\eta>0$ such that $f(x,\epsilon)>\delta$ for $x\in (0,\eta)$, implying $x\leq g_{\phi,\delta}(x)\leq \epsilon$ for $x\in (0,\eta)$. Consequently, $\lim_{x\downarrow 0}g_{\phi,\delta}(x)=0$.

Combining the above conclusions, we obtain the continuity of $g_{\phi,\delta}$ over $[0,1]$. 

In the above, it shows that $f(x,1)$ is strictly decreasing and continuous for $x\in (0,1)$ and $\lim_{x\uparrow 1}f(x,1)=0<\delta$. Note that  $x_\delta\in (0,1)$  satisfies $f(x_\delta,1)=\delta$. Hence, $x_\delta\uparrow 1$ as $\delta\to 0$.
 Therefore, we have for $x\in (0,1)$, $g_{\phi,\delta}(x)>x$ and  $g_{\phi,\delta}(x)\downarrow x$ as $\delta\downarrow 0$. This completes the proof.
\qed

{\bf Proof of Theorem \ref{cor:upper1}}. In light of Proposition \ref{Prop:alpha}, we immediately obtain 
$\sup_{\Q\in \mathcal P_\phi(\p,\delta)}\Lambda\VaR^\Q=\Lambda\VaR^{g_{\phi,\delta}\circ\p}$ and for $\alpha\in (0,1)$, $\sup_{\Q\in \mathcal P_\phi(\p,\delta)}\VaR_{\alpha}^\Q=\VaR_{\alpha}^{g_{\phi,\delta}\circ\p}$.  
Moreover,  if $\phi:[0,\infty)\to\R$ is strictly convex and  $\lim_{x\to\infty}\frac{\phi(x)}{x}=\infty$, then using the properties of $g_{\phi,\delta}$ given in  Proposition \ref{Prop:alpha}, we have 
$\sup_{\Q\in \mathcal P_\phi(\p,\delta)}\Lambda\VaR^\Q=(g_{\phi,\delta}^{-1}\circ\Lambda)\VaR^\p$ and $\sup_{\Q\in \mathcal P_\phi(\p,\delta)}\VaR_{\alpha}^\Q=\VaR_{g_{\phi,\delta}^{-1}(\alpha)}^\p$.
We further have 
$\lim_{\delta\downarrow 0}\sup_{\Q\in \mathcal P_\phi(\p,\delta)}\Lambda\VaR^\Q(X)=\lim_{\delta\downarrow 0}(g_{\phi,\delta}^{-1}\circ\Lambda)\VaR^\p(X).$ Next, we will show $$\lim_{\delta\downarrow 0}(g_{\phi,\delta}^{-1}\circ\Lambda)\VaR^\p(X)=\Lambda\VaR^{+,\p}(X).$$ 
It follows from Proposition \ref{Prop:alpha} that for $x\in (0,1)$, $g_{\phi,\delta}(x)>x$ and $g_{\phi,\delta}(x)\downarrow x$ as $\delta\downarrow 0$, which implies for $t\in (0,1)$, $g_{\phi,\delta}^{-1}(t)<t$ and $g_{\phi,\delta}^{-1}(t)\uparrow t$ as $\delta\downarrow 0$. As $0<\Lambda(x)<1$, then $x_0=\Lambda\VaR^{+,\p}(X)\in \R$. Hence, we have $\p(X>x)<\Lambda(x)$ for $x>x_0$ and 
$\p(X>x)\geq \Lambda(x)$ for $x<x_0$. This implies $\p(X>x)\geq \Lambda(x)>g_{\phi,\delta}^{-1}\circ \Lambda(x) $ for $x<x_0$. By the definition of $\Lambda\VaR$, we have for $\delta>0$,
$(g_{\phi,\delta}^{-1}\circ\Lambda)\VaR^\p(X)\geq x_0=\Lambda\VaR^{+,\p}(X),$
which implies 
$$\liminf_{\delta\downarrow 0}(g_{\phi,\delta}^{-1}\circ\Lambda)\VaR^\p(X)\geq\Lambda\VaR^{+,\p}(X).$$
Moreover, for $x_1>x_0$, using the fact $g_{\phi,\delta}^{-1}(t)\uparrow t$ as $\delta\downarrow 0$ and  $\p(X>x_1)<\Lambda(x_1)$, there exists $\delta_0>0$ such that $\p(X>x_1)<g_{\phi,\delta}^{-1}\circ \Lambda(x_1)$ for $\delta<\delta_0$. Hence, for $\delta<\delta_0$.
$$(g_{\phi,\delta}^{-1}\circ\Lambda)\VaR^\p(X)\leq x_1,$$
which implies  $\limsup_{\delta\downarrow 0}(g_{\phi,\delta}^{-1}\circ\Lambda)\VaR^\p(X)\leq x_1$. By the arbitrariness of $x_1$, we have 
$$\limsup_{\delta\downarrow 0}(g_{\phi,\delta}^{-1}\circ\Lambda)\VaR^\p(X)\leq \Lambda\VaR^{+,\p}(X).$$
Combining the above results, we have $\lim_{\delta\downarrow 0}(g_{\phi,\delta}^{-1}\circ\Lambda)\VaR^\p(X)=\Lambda\VaR^{+,\p}(X).$
This completes the proof. \qed

{\bf Proof of Corollary \ref{cor:upper}}. In light of Proposition \ref{Prop:alpha}, we have 
$$\sup_{\mathbb Q\in \mathcal P_\phi(\p,\delta)}\rho_g^{\Q}(X)\leq \rho_{g\circ g_{\phi,\delta}}^\p(X).$$
By Proposition \ref{Prop:alpha} again, we have  for $x\in [0,1]$,  $g_{\phi,\delta}(x)\downarrow x$ as $\delta\downarrow 0$, which together with the right continuity of $g$ implies $g\circ g_{\phi,\delta}(t)\downarrow g(t)$ as $\delta\downarrow 0$ for $t\in [0,1]$. It follows from the monotone convergence theorem  that $\lim_{\delta\downarrow 0}\rho_{g\circ g_{\phi,\delta}}^\p(X)=\rho_{g}^\p(X)$, which further implies
  $$\lim_{\delta\downarrow 0}\sup_{\mathbb Q\in \mathcal P_\phi(\p,\delta)}\rho_g^{\Q}(X)=\lim_{\delta\downarrow 0}\rho_{g\circ g_{\phi,\delta}}^\p(X)=\rho_{g}^\p(X).$$
This completes the proof. \qed

{\bf Proof of Proposition \ref{thm:likelihood}}. For $A\in\mathcal F$, if  $\mathbb E^{\mathbb P}(Y_2\id_{A}+Y_1\id_{A^c})\leq 1$, then there exists $Y$ satisfying $Y_1\leq Y\leq Y_2$ and $\mathbb E^{\mathbb P}(Y_2\id_{A}+Y\id_{A^c})=1$.  We choose $\Q_0$ with $\frac{d\Q_0}{d\mathbb P}=Y_2\id_{A}+Y\id_{A^c}$. Clearly, $\Q_0\in\mathcal P(\p,Y_1,Y_2)$ and $\Q_0(A)=\mathbb E^{\mathbb P}(Y_2\id_{A})$. Hence, 
    $\overline{\mathbb Q}(A)\geq \mathbb E^{\mathbb P}(Y_2\id_{A})$. The converse inequality holds trivially.  We obtain the desired result.
    
Next, we consider the case  $\mathbb E^{\mathbb P}(Y_2\id_{A}+Y_1\id_{A^c})>1$.  Clearly, there exists  $B\subseteq A$ such that $\mathbb E^{\mathbb P}(Y_2\id_{B}+Y_1\id_{B^c})=1$. For any $\Q\in\mathcal P(\p,Y_1,Y_2)$, we have
\begin{align*}
0&=\mathbb E^{\mathbb P}(Y_2\id_{B}+Y_1\id_{B^c})-\mathbb E^{\mathbb P}\left(\frac{d\Q}{d\p}\id_{B}+\frac{d\Q}{d\p}\id_{B^c}\right)\\
&=\mathbb E^{\mathbb P}\left(\left(Y_2-\frac{d\Q}{d\p}\right)\id_{B}\right)-\mathbb E^{\mathbb P}\left(\left(\frac{d\Q}{d\p}-Y_1\right)\id_{B^c}\right)\\
&\leq \mathbb E^{\mathbb P}\left(\left(Y_2-\frac{d\Q}{d\p}\right)\id_{B}\right)-\mathbb E^{\mathbb P}\left(\left(\frac{d\Q}{d\p}-Y_1\right)\id_{A\setminus B}\right),
\end{align*}
which implies $$\Q(A)=\mathbb E^{\mathbb P}\left(\frac{d\Q}{d\p}\id_{A}\right)\leq \mathbb E^{\mathbb P}(Y_2\id_{B}+Y_1\id_{A\setminus B}).$$
Hence, we have 
    $$\overline{\mathbb Q}(A)\leq \mathbb E^{\mathbb P}(Y_2\id_{B}+Y_1\id_{A\setminus B}).$$
    Moreover, we choose $\Q_0$ with $\frac{d\Q_0}{d\mathbb P}=Y_2\id_{B}+Y_1\id_{B^c}$. Clearly, $\Q_0\in\mathcal P(\p,Y_1,Y_2)$ and $\Q_0(A)=\mathbb E^{\mathbb P}(Y_2\id_{B}+Y_1\id_{A\setminus B})$, which implies  $$\overline{\mathbb Q}(A)\geq \mathbb E^{\mathbb P}(Y_2\id_{B}+Y_1\id_{A\setminus B}).$$
     Hence, if $\mathbb E^{\mathbb P}(Y_2\id_{A}+Y_1\id_{A^c})>1$, we have $\overline{\mathbb Q}(A)=\mathbb E^{\mathbb P}(Y_2\id_{B}+Y_1\id_{A\setminus B}).$

{\bf Proof of Proposition \ref{Prop:likelihood1}}.
 Case (i) is obvious. We next focus on case (ii). If $\p(A)\leq \frac{1-k_{1}}{k_{2}-k_{1}}$, then $\mathbb E^{\mathbb P}(k_2\id_{A}+k_1\id_{A^c})\leq 1$.  It follows from Proposition \ref{thm:likelihood} that $\overline{\Q}(A)= k_{2}\p(A)$.

If $\p(A)>\frac{1-k_{1}}{k_{2}-k_{1}}$,  then $\mathbb E^{\mathbb P}(k_2\id_{A}+k_1\id_{A^c})>1$.  We choose $B\subseteq A$ such that $\p(B)=\frac{1-k_{1}}{k_{2}-k_{1}}$. Then $\mathbb E^{\p}(k_{2}\id_{B}+k_{1}\id_{B^c})=1$. In light of Proposition \ref{thm:likelihood}, we have  $\overline{\Q}(A)=\mathbb E^{\p}(k_{2}\id_{B}+k_{1}\id_{A\setminus B})=g(\p(A))$ with $g(x)=(k_{2}x)\wedge (k_{1}x+1-k_{1}),~x\in [0,1]$.  This completes the proof. \qed

\section{Proof of Section \ref{sec:multiple}.}
In this section, we provide the proofs of all the results in Section \ref{sec:multiple}.

{\bf Proof of Theorem \ref{Th:sharing}}.  We first show that $\dsquare_{i=1}^n\rho_i(X)\geq \inf\{x\in\R: \{X>x\}\in \widetilde{\mathcal A}_{x}\}$. For $(X_1,\dots, X_n)\in \mathbb A_n(X)$, let $x_0=\sum_{i=1}^n\rho_i(X_i)$ and $y_i=\rho_i(X_i)\in [-\infty,\infty)$.  By the definition of $\rho_i$, there exists a family $\{y_{i,m}\}_{m\geq 1}$ such that $y_{i,m}>-\infty$, $y_{i,m}\downarrow y_i$ as $m\to\infty$, and $\{X_i>y_{i,m}\}\in\mathcal A_{y_{i,m}}^{(i)}$. Let $x_m=\sum_{i=1}^n y_{i,m}$.  Then it follows that
$$\left\{X>x_m\right\}=\left\{\sum_{i=1}^{n}X_i>x_m\right\}\subseteq \bigcup_{i=1}^n \{X_i>y_{i,m}\}.$$ Define
$$A_{i,m}=\left\{X>x_m\right\}\cap \{X_i>y_{i,m}\}.$$
Then we have
$$\left\{X>x_m\right\}=\cup_{i=1}^n A_{i,m},~A_{i,m}\subseteq\{X_i>y_{i,m}\}.$$
Using the fact that $\mathcal A_{y_{i,m}}^{(i)}$ is a {downset}, we have $A_{i,m}\in\mathcal A_{y_{i,m}}^{(i)}$. Hence, 
 $\left\{X>x_m\right\}\in \widetilde{\mathcal A}_{x_m}$, which implies
  $\inf\{x\in\R: \{X>x\}\in \widetilde{\mathcal A}_{x}\}\leq x_m$. Letting $m\to\infty$, we have $$\inf\{x\in\R: \{X>x\}\in \widetilde{\mathcal A}_{x}\}\leq x_0=\sum_{i=1}^{n}\rho_i(X_i).$$ Using the arbitrariness of $(X_1,\dots, X_n)\in \mathbb A_n(X)$, we have $$\inf\{x\in\R: \{X>x\}\in \widetilde{\mathcal A}_{x}\}\leq \dsquare_{i=1}^n\rho_i(X).$$

Next we show the inverse inequality. Suppose  $\{X>x\}\in \widetilde{\mathcal A}_{x}$. Then there exist $y_1,\dots, y_n\in\R$ and $B_1,\dots B_n\in\mathcal F$ such that $\sum_{i=1}^{n}y_i=x$, $\cup_{i=1}^n B_i=\{X>x\}$ and $B_i\in\mathcal A_{y_i}^{(i)}$. This implies there exists  $(A_1,\dots, A_n)\in\Pi_n(\Omega)$ such that $\{X>x\}\cap A_i\subseteq B_i$. Note that $\mathcal A_{y_i}^{(i)}$ is a {downset}. Hence, $\{X>x\}\cap A_i\in\mathcal A_{y_i}^{(i)}$.  Let $X_i=(X-x)\id_{A_i}+y_i,~i\in [n]$. Then $(X_1,\dots, X_n)\in \mathbb A_n(X)$. It follows that  $$\{X_i>y_i\}=\{X>x\}\cap A_i\in\mathcal A_{y_i}^{(i)},$$ which implies $\rho_i(X_i)\leq y_i$. Hence,
$$\sum_{i=1}^{n}\rho_i(X_i)\leq \sum_{i=1}^{n} y_i=x.$$
This implies $\dsquare_{i=1}^n\rho_i(X)\leq  x$. By the arbitrariness of $x$, we have $\dsquare_{i=1}^n\rho_i(X)\leq \inf\{x\in\R: \{X>x\}\in \widetilde{\mathcal A}_{x}\}$. Combing the above results, we have $\dsquare_{i=1}^n\rho_i(X)=\inf\{x\in\R: \{X>x\}\in \widetilde{\mathcal A}_{x}\}$.

One can directly check that $(X_1^*,\dots,X_n^*)$ is the optimal allocation.

If $\{\mathcal A_x^{(i)}\}_{x\in\R},~i\in [n]$, are all increasing, then we have
$$\widetilde{\mathcal A}_x=\bigcup_{y_1+\dots+y_n\leq x}\oplus_{i=1}^n\mathcal A_{y_i}^{(i)},$$
implying that $\{\widetilde{\mathcal A}_x\}_{x\in\R}$ is increasing.

If $\{\mathcal A_x^{(i)}\}_{x\in\R},~i\in [n]$, are all decreasing, then we have
$$\widetilde{\mathcal A}_x=\bigcup_{y_1+\dots+y_n\geq x}\oplus_{i=1}^n\mathcal A_{y_i}^{(i)},$$
which implies that $\{\widetilde{\mathcal A}_x\}_{x\in\R}$ is decreasing. Moreover, direct computation gives
$$\bigcap_{y<x}\widetilde{\mathcal A}_{y}=\bigcap_{y<x}\bigcup_{y_1+\dots+y_n\geq y}\oplus_{i=1}^n\mathcal A_{y_i}^{(i)}=\bigcup_{y_1+\dots+y_n\geq x}\oplus_{i=1}^n\mathcal A_{y_i}^{(i)}=\widetilde{\mathcal A}_{x}.$$
Hence, $\{\widetilde{\mathcal A}_x\}_{x\in\R}$ is left-continuous.
 This completes the proof. \qed

{\bf Proof of Proposition \ref{Thmain}}.
Note that $\rho_i=\Lambda_i\VaR^{w_i}$ is induced by the family of downsets below: \begin{align*}
    \mathcal A_{x}^{(i)}=\left\{A\in\mathcal F: w_i(A)\leq \Lambda_i(x)\right\},~x\in\R.
\end{align*}
Hence $\widetilde{\mathcal A}_{x}=\mathcal A_{x}^{\mathbf \Lambda, \mathbf w}$. It follows from the fact that $\lambda_i^->0$ and the continuity of $w_i$ that $\Lambda_i\VaR^{w_i}(X)\in [-\infty,\infty)$ for all $X\in\mathcal X$.  Then the expressions of the inf-convolution and the optimal allocation follow directly from Theorem \ref{Th:sharing}.

Moreover, if all $\Lambda_i$ are increasing, then $\{\mathcal A_{x}^{(i)}\}_{x\in\R}, i\in [n]$ are increasing. Applying (i) of Theorem \ref{Pro:representation}, we have  $X\mapsto\inf\{x\in\R: \{X>x\}\in \mathcal A_{x}^{\mathbf \Lambda, \mathbf w}\}$ is a $\Lambda\VaR^w$ with $\Lambda\in\mathcal H_I^*$.
If all $\Lambda_i$ are decreasing,  then all $\{\mathcal A_{x}^{(i)}\}_{x\in\R}$ are decreasing. It follows from Theorem \ref{Th:sharing} that $\{\mathcal A_{x}^{\mathbf \Lambda, \mathbf w}\}_{x\in\R}$ is decreasing and left-continuous.  Applying (ii) of Theorem \ref{Pro:representation},   $X\mapsto\inf\{x\in\R: \{X>x\}\in \mathcal A_{x}^{\mathbf \Lambda, \mathbf w}\}$ is a $\Lambda\VaR^w$ with $\Lambda\in\mathcal H_D^*$. \qed

{\bf Proof of Proposition \ref{prop:finite2}}. Note that $\Lambda_i\in\mathcal H_I$. If $\bigwedge_{i=1}^n\left(\frac{w_i(A_i)}{\lambda_i^-}\vee\bigvee_{j\neq i}\frac{w_j(A_j)}{\lambda_j^+}\right)<1$ for some $(A_1,\dots, A_n)\in\Pi_n(\Omega)$, then there exists $i\in\{1,\dots,n\}$ such that
$w_i(A_i)<\lambda_i^-$ and $w_j(A_j)<\lambda_j^+$ for $j\neq i$. Note that there exist $y_j^0\in\R$ such that $w_j(A_j)\leq \Lambda_j(y_{j}^0)$ for all $j\neq i$. Moreover, for any $y_i\in\R$, it follows that $w_j(\{X>y_i+\sum_{k\neq i} y_k^0\}\cap A_j)\leq w_j(A_j)\leq \Lambda_j(y_j^0)$ for $j\neq i$, and $w_i(\{X>y_i+\sum_{k\neq i} y_k^0\}\cap A_i)\leq \Lambda_i(y_i)$. This implies 
$$\left\{X>y_i+\sum_{k\neq i} y_k^0\right\}\in \mathcal A_{y_i+\sum_{k\neq i} y_k^0}^{\mathbf \Lambda, \mathbf w}.$$
Hence, $\Gamma_{\mathbf \Lambda, \mathbf w}(X)\leq y_i+\sum_{k\neq i} y_k^0$.
By the arbitrariness of $y_i$, we have $\Gamma_{\mathbf \Lambda, \mathbf w}(X)=-\infty$.

We next show the second statement. 
Let $1+\epsilon=\inf_{(A_1,\dots, A_n)\in\Pi_n(\Omega)}\bigwedge_{i=1}^n\left(\frac{w_i(A_i)}{\lambda_i^-}\vee\bigvee_{j\neq i}\frac{w_j(A_j)}{\lambda_j^+}\right)$ for some $\epsilon>0$.
Suppose by contradiction that $\Gamma_{\mathbf \Lambda, \mathbf w}(X)=-\infty$. Then there exist $(y_1^{(m)},\cdots, y_n^{(m)})$ satisfying $\lim_{m\to\infty}\sum_{i=1}^{n}y_i^{(m)}=-\infty$, and  $(A_1^{(m)},\dots, A_n^{(m)})\in\Pi_n(\Omega)$ such that $$w_i\left(\left\{X>\sum_{j=1}^n y_j^{(m)}\right\}\cap A_i^{(m)}\right)\leq \Lambda_i(y_i^{(m)}).$$ Without loss of generality, we can assume  $\lim_{m\to\infty}y_i^{(m)}=y_i$. Let $E_1=\{i\in\{1,\dots, n\}: y_i=-\infty\}$ and $E_2=\{i\in\{1,\dots, n\}: y_i>-\infty\}$.   By the fact that $\lim_{m\to\infty}\sum_{i=1}^{n}y_i^{(m)}=-\infty$, we have $E_1\neq\emptyset$.  Moreover, for any $0<\eta<\epsilon$, using the continuity of $w_i$, 
there exists $m_0>0$ such that $w_i(X\leq\sum_{j=1}^n y_j^{(m)})<\lambda_i^-\eta/2$ for all $m\geq m_0$. It follows from the subadditivity of $w_i$ that $$w_i(A_i^{(m)})\leq w_i\left(X\leq\sum_{j=1}^n y_j^{(m)}\right)+w_i\left(\left\{X>\sum_{j=1}^n y_j^{(m)}\right\}\cap A_i^{(m)}\right)\leq (1+\eta/2)\Lambda_i(y_i^{(m)})$$ for all $m\geq m_0$. Note that $\lim_{m\to\infty}\Lambda_i(y_i^{(m)})=\lambda_i^-$ for all $i\in E_1$. Hence, for any $0<\eta<\epsilon$,  there exists $m_1\geq m_0$ such that $w_i(A_i^{(m)})\leq \lambda_i^-(1+\eta)$ for all $m\geq m_1$ and $i\in E_1$. Moreover, $w_i(A_i^{(m)})\leq \lambda_i^+(1+\eta)$ for all $m\geq m_1$ and $i\in E_2$. Combination of the those conclusion leads to $\bigwedge_{i=1}^n\left(\frac{w_i(A_i^{(m)})}{\lambda_i^-}\vee\bigvee_{j\neq i}\frac{w_j(A_j^{(m)})}{\lambda_j^+}\right)\leq 1+\eta<1+\epsilon$ for $m>m_1$, which is a contradiction of the assumption. Hence, $\Gamma_{\mathbf \Lambda, \mathbf w}(X)>-\infty$. We complete the proof.

 {\bf Proof of Corollary \ref{prop:likelihood}}. Using Proposition \ref{Th:likelihood}, the attainability of $\overline{\Lambda}^*$ and the atomlessness of $\p$, we have
\begin{align*}
&\dsquare_{i=1}^n\sup_{\mathbb Q\in \mathcal P(\p,0,Y)}\Lambda_i\VaR^{\mathbb Q}(X)\\
&=\inf\{x\in\R: \{X>x\}\in \mathcal A_{x}^{\mathbf \Lambda, (\overline {\mathbb Q}_1,\dots, \overline {\mathbb Q}_n)}\}\\
&=\inf\left\{x\in\R: 1\wedge\mathbb E^{\mathbb P}\left(Y\id_{{\{X>x\}}\cap A_i}\right)\leq \Lambda_i(y_i)~\text{for some}~\sum_{j=1}^{n}y_j=x, (A_1,\dots, A_n)\in\Pi_n(\Omega)\right\}\\
&=\inf\left\{x\in\R: \mathbb E^{\mathbb P}\left(Y\id_{{\{X>x\}}}\right)\leq \sum_{i=1}^n\Lambda_i(y_i)~\text{for some}~\sum_{j=1}^{n}y_j=x\right\}\\
&=\inf\left\{x\in\R: \mathbb E^{\mathbb P}\left(Y\id_{{\{X>x\}}}\right)\leq \overline{\Lambda}^*(x)\right\}.
\end{align*}

The right continuity of $\overline{\Lambda}^*$ at $x^*$ implies $\mathbb E^{\mathbb P}\left(Y\id_{{\{X>x^*\}}}\right)\leq \overline{\Lambda}^*(x^*)$. The attainability of $\overline{\Lambda}^*$ implies the existence of $(y_1^*,\dots,y_n^*)$ such that $\sum_{i=1}^ny_i^*=x^*$ and $\sum_{i=1}^n\Lambda_i(y_i^*)=\overline{\Lambda}^*(x^*)$. Hence, we have $\mathbb E^{\mathbb P}\left(Y\id_{{\{X>x^*\}}}\right)\leq \sum_{i=1}^n\Lambda_i(y_i^*)$. Then we can choose $(A_1^*,\dots, A_n^*)\in \mathbb A_n(X)$ such that $\mathbb E^{\mathbb P}\left(Y\id_{\{X>x^*\}\cap A_i^*}\right)\leq \Lambda_i(y_i^*)$. Hence, $x^*$ is the minimizer of $\inf\{x\in\R: \{X>x\}\in \mathcal A_{x}^{\mathbf \Lambda, (\overline {\mathbb Q}_1,\dots, \overline {\mathbb Q}_n)}\}$ in Proposition \ref{Th:likelihood}, and the optimal allocation is given by \eqref{Optimal}. We complete the proof. \qed

{\bf Proof of Corollary \ref{Prop:111}}.  In light of  Proposition \ref{Th:likelihood}, we have
\begin{align*}
&\dsquare_{i=1}^n\sup_{\mathbb Q\in \mathcal P(\p,0,Y)}\Lambda_i\VaR^{\mathbb Q}(X)\\
&=\inf\left\{x\in\R: 1\wedge\mathbb E^{\mathbb P}\left(Y\id_{{\{X>x\}}\cap A_i}\right)\leq \Lambda_i(y_i)~\text{for some}~\sum_{j=1}^{n}y_j=x, (A_1,\dots, A_n)\in\Pi_n(\Omega)\right\}\\
&=\inf\left\{x\in\R: \mathbb E^{\mathbb P}\left(Y\id_{{\{X>x\}}}\right)\leq \sum_{i=1}^n\Lambda_i(y_i)~\text{for some}~\sum_{j=1}^{n}y_j=x\right\}\\
&=\inf_{\mathbf y_{n-1}\in\R^{n-1}}\inf\left\{x\in\R: \mathbb E^{\mathbb P}\left(Y\id_{{\{X>x\}}}\right)\leq \Lambda^{\mathbf{y}_{n-1}}(x)\right\}.
\end{align*}
\section{Proof of Section \ref{Sec:Comonotonic}}
We give the proof of the result in Section \ref{Sec:Comonotonic} below.

{\bf Proof of Theorem \ref{Thmain1}}.
Let $$\mathbb F_n=\{(f_1,\dots,f_n): f_1+\dots+f_n=id,~\text{all}~f_i~\text{are increasing and continuous and }f_i(0)=0\},$$ where $id$ is the {identity} function.
Using the comonotonic assumption \trd{and Proposition 4.5 of \cite{D94}}, we have
\begin{align*}
    \boxplus_{i=1}^n\Lambda_i\VaR^{w_i}(X)=\inf\left\{\sum_{i=1}^n\Lambda_i\VaR^{w_i}(f_i(X)): (f_1,\dots,f_n)\in \mathbb F_n  \right\}.
\end{align*}
    It follows from the definition that
    \begin{align*}
    \Lambda_i\VaR^{w_i}(f_i(X))&=\inf\{x\in\R: w_i(f_i(X)>x)\leq \Lambda_i(x)\}\\
    &=\inf\{x\in\R: w_i(X>f_i^{-1,+}(x))\leq \Lambda_i(x)\}\\
    &=\inf\{f_i(x)\in\R: w_i(X>x)\leq \Lambda_i(f_i(x))\}\\
    &=f_i((\Lambda_i\circ f_i)\VaR^{w_i}(X)),
    \end{align*}
    where $f_i^{-1,+}(t)=\inf\{x\in \R: f_i(x)>t\}$ for $t\in \R$ with the convention that $\inf\emptyset=\infty$.
    Note that $0\leq f_i(x)\leq x$ if $x\geq 0$ and $x\leq f_i(x)\leq 0$ if $x\leq 0$, which implies $\Lambda_i(f_i(x))\leq \Lambda_i(x)$ for $x\geq 0$  and  $\Lambda_i(f_i(x))\geq \Lambda_i(x)$ for $x\leq 0$.
Hence, if $x_i=\Lambda_i\VaR^{w_i}(X)>0$, we have  $w_i(X>x)>\Lambda_i(x)\geq \Lambda_i(f_i(x))$ for $0<x<x_i$, which implies $(\Lambda_i\circ f_i)\VaR^{w_i}(X)\geq x_i$. 

Next, we consider the case $x_i=0$. Then we have  $w_i(X>x)>\Lambda_i(x)$ if $x<0$. If $f_i(x)<0$ for all $x<0$, then   $w_i(X>x)\geq w_i(X>f_i(x))>\Lambda_i(f_i(x))$ for all $x<0$, implying $(\Lambda_i\circ f_i)\VaR^{w_i}(X)\geq 0$. If  $f_i(x)=0$ for some $x<0$, then   let $y_i=\inf\{x\in\R: f_i(x)=0\}$. Then, $f_i(x)<0$ for all $x<y_i$. This implies $w_i(X>x)\geq w_i(X>f_i(x))>\Lambda_i(f_i(x))$ for all $x<y_i$. Hence, $(\Lambda_i\circ f_i)\VaR^{w_i}(X)\geq y_i$.

Combining all the above conclusions, we have \begin{align}\label{Eq:comotonicproof}\Lambda_i\VaR^{w_i}(f_i(X))=f_i((\Lambda_i\circ f_i)\VaR^{w_i}(X))\geq f_i(\Lambda_i\VaR^{w_i}(X)).
\end{align}
Then we further have
\begin{align}\label{Eq:comonotonic1}\sum_{i=1}^n\Lambda_i\VaR^{w_i}(f_i(X))\geq \sum_{i=1}^nf_i(\Lambda_i\VaR^{w_i}(X))\geq \min_{i \in [n]} \Lambda_i\VaR^{w_i}(X).\end{align}

By choosing $f_{i_0}=id$ and $f_i=0$ for $i\neq i_0$ with $ i_0=\argmin_{i \in [n]}\Lambda_i\VaR^{w_i}(X)$, we have $\sum_{i=1}^n\Lambda_i\VaR^{w_i}(f_i(X))=\min_{i \in [n]} \Lambda_i\VaR^{w_i}(X)$, which implies our desired result.

Note that if $\Lambda_i,~i=1\dots, n$, are all constants over $(-\infty,0)$, then $\Lambda_i\circ f_i\leq \Lambda_i$ for $(f_1,\dots,f_n)\in \mathbb F_n$, which implies $(\Lambda_i\circ f_i)\VaR^{w_i}(X)\geq \Lambda_i\VaR^{w_i}(X)$. Hence,
\eqref{Eq:comotonicproof} and \eqref{Eq:comonotonic1} are valid for all $X\in \mathcal X$, which further implies \eqref{eq:multiple11}.

Finally, we show that this condition is also necessary  for \eqref{eq:multiple11} if all $w_1=\dots=w_n$ are atomless and $\Lambda_1=\dots=\Lambda_n$.  Assume by contradiction that $\Lambda_1$ is not a constant over $(-\infty,0)$.  Then there exist two continuous points of $\Lambda_1$, denoted by  $x_1,x_2$,  satisfying $x_1<x_2<0$ and $0<\Lambda_1(x_1)<\Lambda_1(x_2)<1$. The atomlessness of $w_1$ means that there exists a bounded random variable $Y\in\X$ such that $w_1(Y>x)$ is continuous over $\R$. Let $h(x)=1-w_1(Y>x)$. Then we have $w_1(h(Y)>t)=1-t$ for $t\in (0,1)$.   
Let $X=\frac{-x_1}{\Lambda_1(x_1)}(h(Y)-1)$. 
Direct computation shows $$w_1(X>x)=w_1(h(Y)>1-(x\Lambda_1(x_1))/x_1),~x\in\R.$$
Hence, $w_1(X>x_1)=\Lambda_1(x_1)$ and $w_1(X>x)<\Lambda_1(x_1)$ if $x>x_1$ and $w_1(X>x)>\Lambda_1(x_1)$ if $x<x_1$. This implies 
$\Lambda_1\VaR^{w_1}(X)=x_1$.   
 For $(f_1,\dots,f_n)\in \mathbb F_n$, let  $f_1(x)=\frac{\mid x_2\mid}{\mid x_1\mid }x$.  It follows that  $w_1(X>x_1)=\Lambda_1(x_1)<\Lambda_1(x_2)=\Lambda_1\circ f_1(x_1)$. Using the fact that $w_1(X>x)$ and $\Lambda_1\circ f_1$ are continuous at $x_1$, we could find $\delta>0$ such that $w_1(X>x)<\Lambda_1\circ f_1(x)$ holds over $(x_1-\delta,x_1)$. By the definition, we have $(\Lambda_1\circ f_1)\VaR^{w_1}(X)\leq \Lambda_1\VaR^{w_1}(X)-\delta$, which implies $f_1((\Lambda_1\circ f_1)\VaR^{w_1}(X))<f_1(\Lambda_1\VaR^{w_1}(X))$.

 Recall that $w_1=\dots=w_n$ and $\Lambda_1=\dots=\Lambda_n>0$. It follows from the fact $w_1(X>0)=0$ and the continuity of $w_1(X>x)$ that $w_i(X>\epsilon)\leq \Lambda_i(\epsilon)$ for some $\epsilon<0$. Hence, $\Lambda_i\VaR^{w_i}(X)\leq \epsilon<0$, which together with the fact that $\Lambda_1\circ f_i(x)\geq \Lambda_1(x)$ over $(-\infty, 0]$ further implies $(\Lambda_i\circ f_i)\VaR^{w_i}(X)\leq \Lambda_i\VaR^{w_i}(X)$.
Consequently, noting that $w_1=\dots=w_n$ and $\Lambda_1=\dots=\Lambda_n$, we have
\begin{align*}\sum_{i=1}^n\Lambda_i\VaR^{w_i}(f_i(X))=\sum_{i=1}^n f_i((\Lambda_i\circ f_i)\VaR^{w_i}(X))<\sum_{i=1}^nf_i(\Lambda_i\VaR^{w_i}(X))=\min_{i \in [n]} \Lambda_i\VaR^{w_i}(X),\end{align*}
which contradicts \eqref{eq:multiple11}. We complete the proof.  \qed

\end{document}